\documentclass[a4paper,submit,oneauthors]{article}

\usepackage{color}
\usepackage{soul}
\usepackage{natbib}
\usepackage{graphicx}
\usepackage {changepage}
\usepackage{booktabs}
\usepackage{subfig}

\newcommand{\be}{\begin{equation}}
\newcommand{\ee}{\end{equation}}
\newcommand{\bra}{\langle}
\newcommand{\ket}{\rangle}
\newcommand{\bea}{\begin{eqnarray}}
\newcommand{\eea}{\end{eqnarray}}

\begin{document}
\title{\Large Impact of the COVID-19 pandemic on the financial market efficiency of price returns, absolute returns, and volatility increment: Evidence from stock and cryptocurrency markets}

\author{Tetsuya Takaishi \\
\small{Hiroshima University of Economics, Hiroshima 731-0192, Japan} \\
\small{tt-taka@hue.ac.jp}
} 


\maketitle

\abstract{
This study examines the impact of the coronavirus disease 2019 (COVID-19) pandemic on market efficiency by analyzing three time series---price returns, absolute returns, and volatility increments---in stock (Deutscher Aktienindex, Nikkei 225, Shanghai Stock Exchange (SSE), and Volatility Index) and cryptocurrency (Bitcoin and Ethereum) markets. The effect is found to vary by asset class and market.
In the stock market, while the pandemic did not influence the Hurst exponent of volatility increments, it affected that of returns and absolute returns (except in the SSE, where returns remained unaffected). In the cryptocurrency market, the pandemic did not alter the Hurst exponent for any time series but influenced the strength of multifractality in returns and absolute returns.
Some Hurst exponent time series exhibited a gradual decline over time, complicating the assessment of pandemic-related effects. Consequently, segmented analyses by pandemic periods may erroneously suggest an impact, warranting caution in period-based studies.
}

Keywords: market efficiency; generalized Hurst exponent; 
multifractality; COVID-19; rough volatility; finite-sample effect

\section{Introduction}

\citet{Fama1970efficient} developed and classified the efficient market hypothesis (EMH) into three forms: (i) weak, (ii) semi-strong, and (iii) strong. 
The weak-form efficient market considers only historical prices, asserting that current prices include all past information, making future prices unpredictable.
Weak-form efficient markets have been extensively tested.
To date, no definitive answer has been obtained. Rather, the possibility that the market efficiency may change over time 
was discussed in \citep{lim2011evolution}.

One statistical model that can match this market is the random walk model.
Therefore, numerous studies have attempted to determine whether time series exhibit randomness.
A popular method for testing the randomness of a time series is to measure its Hurst exponent $h$ of the time series, introduced by Hurst\citep{hurst1951long}.
For a random time series, $h=0.5$. Thus, $h\neq 0.5$ is used as an indicator of market inefficiency.
For the time series with $h>0.5$, successive movements in the same direction were observed more often than in the random walk process.
The time series with $h>0.5$ is denoted as "persistent." 
Conversely, time series with $h<0.5$ have successive movement back and forth more often than the random walk process, denoted as "anti-persistent."

\citet{MATTEO2005827} investigated the Hurst exponents of 32 world stock market indices and 
classified them based on their Hurst exponents.
They found that while all emerging markets belong to a group with $h > 0.5$ (persistent), developed markets fall into either groups with $h=0.5$ (random) or $h < 0.5$ (anti-persistent).
Well-developed markets (the USA, Japan, France, and Australia) are classified as follows:
a group with $h < 0.5$, which indicates inefficiency.
\citet{takaishi2022time} investigated the time evolution of the Hurst exponent for the Japanese stock market
and found that in the early stage of the market, the Hurst exponent was higher than 0.5,
gradually decreased and dropped below 0.5 around the year 2000, indicating anti-persistency. 
The Bitcoin market at the early stage also shows anti-persistency for return time series\citep{urquhart2016inefficiency,takaishi2018}.
This anti-persistency gradually disappears as the Bitcoin market matures\citep{drozdz2018bitcoin}. 
The anti-persistency observed in the early stage of the Bitcoin market is linked to 
poor liquidity of the market\citep{wei2018liquidity,takaishi2019market}.

Another anti-persistent time series is found in the volatility increment time series, which
is an important property for modeling and forecasting volatility. 
\citet{gatheral2018volatility} investigated the Hurst exponent of the realized volatility time series and found that the realized volatility time series exhibit anti-persistency, 
denoted as "rough volatility." This property has been further confirmed through subsequent studies on various assets and implied volatility\citep{bennedsen2022decoupling,livieri2018rough,floc2022roughness,takaishi2020rough,brandi2022multiscaling}. 
\citet{takaishi2022tradingvolume} investigated the Hurst exponent of trading volume increment time series and showed that the trading volume increment time series is anti-persistent, and the time variation of the Hurst exponent
is closely related to that of volatility increment time series.

The coronavirus disease 2019 (COVID-19) pandemic, officially declared a pandemic by World Health Organization (WHO) on March 11, 2020, 
significantly affected the global economy. 
Financial markets were also heavily disrupted by COVID-19, suggesting that
market efficiency was affected. The impact of COVID-19 on market efficiency has been studied across various financial asset time series. The results indicate that, in many cases, market efficiency has deteriorated by the impact of COVID-19 
(see, e.g., \citet{aslam2020efficiency,fernandes2022evaluating,wang2020analysis,shen2022multifractal,mensi2021does,alijani2021fractal,zitis2023investigating,raza2024multifractal,erer2023aggregate,aslam2020evidence,mensi2020impact}).

However, COVID-19 seems to affect financial assets differently.
\citet{saadaoui2023skewed} analyzed the impact of COVID-19 on stock market indices and 
found that while the multifractality for the USA, Japanese, and Eurozone markets increased
after the pandemic, the Chinese market’s multifractality decreased.
\citet{ameer2023impact} examined the response of the BRICS (Brazil, Russia, India, China, and South Africa) and MSCI (Morgan Stanley Capital International) emerging stock market indices to the COVID-19 pandemic and found that COVID-19 increased market inefficiency except in China. They concluded that China's market efficiency improved after 
the COVID-19 outbreak.

\citet{mnif2020cryptocurrency} studied the cryptocurrency efficiency and found that 
COVID-19 has positively impacted the cryptocurrency market, and cryptocurrencies have become more
efficient after the spread of the pandemic.
\citet{wang2021covid} examined market efficiency using entropy-based analysis for S\&P 500, gold, Bitcoin, and the US dollar index and found that while these market efficiencies decreased because of COVID-19, 
the Bitcoin market efficiency was more resilient than the others.
\citet{diniz2021bitcoin} also claimed that the Bitcoin prices are more efficient than the US dollar and MSCI world indices.
Conversely, \citet{naeem2021asymmetric} showed that the COVID-19 outbreak adversely affected the efficiency of four cryptocurrencies (Bitocoin, Ethereum, Litecoin and Ripple) given a substantial increase in the levels of inefficiency during the COVID-19 period.

Most previous studies have utilized price-return data for multifractal analysis; however, some studies have examined the impact of COVID-19 using other types of time-series data.
For example, 
using a wavelet-based method, \citet{arouxet2022covid} estimated the Hurst exponent of a volatility time series 
and detected a temporary impact on volatility during the peak of the COVID-19 pandemic.
\citet{lahmiri2023effect}
investigated multifractality for both price returns and trading volume variation time series of cryptocurrencies
and found that the multifractality decreased during the pandemic for both time series.


We investigated the time evolution of market efficiency and examine the impact of the COVID-19 pandemic on the time variation in market efficiency. We analyzed three types of time series---namely, returns, absolute returns, and volatility increment---in order to clarify the effect of the COVID-19 pandemic on these time series. 
To the best of our knowledge, this study is the first to examine the impact of the COVID-19 pandemic on volatility increment.
Recent research on volatility has shown that volatility increment time series provide crucial information for modeling volatility.
After \citet{gatheral2018volatility} demonstrated the roughness of volatility increment time series, modeling volatility using models with roughness became increasingly popular. Several advantages of models with roughness based on fractional Brownian motion are also known. 
For example, it has been pointed out that models with roughness can explain the negative power law of implied volatility\citep{alos2007short,fukasawa2011asymptotic}.
It has also been suggested that rough models could serve as a theoretical framework for explaining the time-reversal asymmetry observed in the correlation between realized volatility and returns---a phenomenon commonly known as the Zumbach effect \citep{zumbach2003volatility,zumbach2009time}. Consequently, analyzing various types of time series, including volatility increment time series, is expected to provide deeper insights into the pandemic's impact on individual markets.

We quantified market efficiency using the generalized Hurst exponent (GHE) obtained via multifractal detrended fluctuation analysis (MFDFA) \citep{kantelhardt2002multifractal} and examined the effect of the COVID-19 pandemic on
the stock and cryptocurrency markets over time evolution of market efficiency.

The remainder of this paper is organized as follows. Section~\ref{sec.2} describes the data used in this study.
Section~\ref{sec.3} describes the proposed methodology.
Section~\ref{sec.4} presents our results. Finally, Section~\ref{sec.5} presents discussion and conclusion.


\section{Data}\label{sec.2}

We retrieved the daily closing prices $P(t)$ of the four indices (DAX, Nikkei 225, SSE, and VIX) 
from https://finance.yahoo.com.
These time-series data cover the period from July 3, 1997, to January 19, 2024, including 
the time of the financial crisis initiated by Lehman and the COVID-19 pandemic.
We also obtained the daily closing prices $P(t)$ of cryptocurrencies (Bitcoin and Ethereum)
from https://www.investing.com. The data collection period was from March 11, 2016, to December 31, 2024, which included the COVID-19 pandemic period only.

The return $r(t)$ is defined as the logarithmic price difference as follows:
\be
r(t)= \log P(t)-\log P(t-1).
\ee 
We use the absolute return $AR(t) \equiv |r(t)|$ as a proxy for volatility,
with a long memory property\citep{ding1993long}.
We also define the volatility increment $VI(t)$ using the absolute return $AR(t)$ as follows:
\be 
VI(t) = \log AR(t)- \log AR(t-1).
\ee
Fig. 1 illustrates the return, absolute return, and volatility increment time series 
of Bitcoin.


\begin{figure}
\includegraphics[height=8cm]{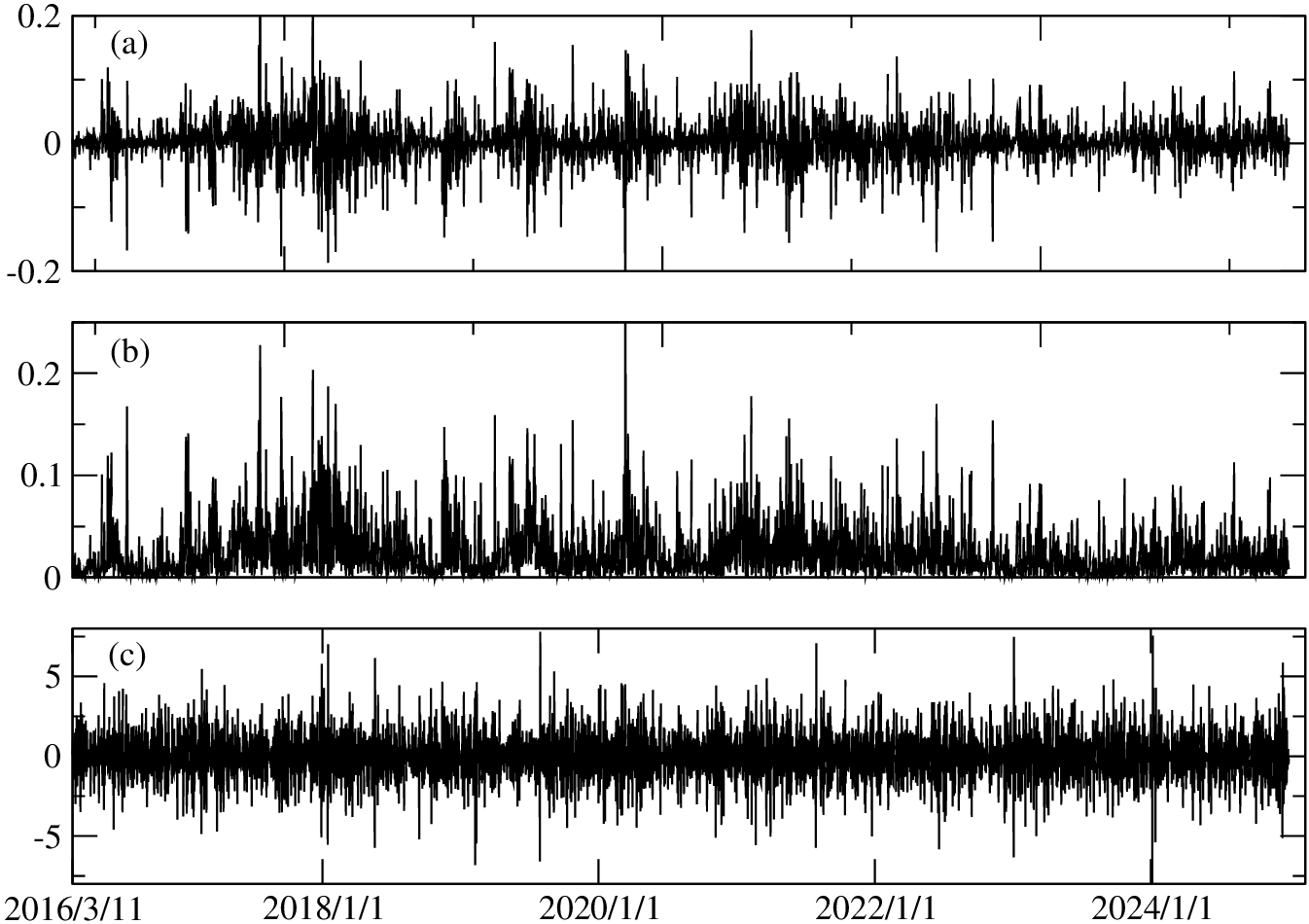}
\caption{Time series of (a) returns, (b) absolute returns, and (c) volatility increment.
\label{fig1}}
\end{figure}

Table 1 summarizes the descriptive statistics of the time series.
The means of most time series for the return and volatility increments are consistent with zero within a one-sigma error.
The kurtosis values of the volatility increment are slightly higher than those of the Gaussian distribution, 
similar to those obtained for realized volatility increments\citep{takaishi2020rough}.

\begin{table}
\centering
\caption{Descriptive statistics of (a) returns, (b) absolute returns, and (c) volatility increment.
The values in the parentheses are statistical errors estimated using the Jackknife method.}
\label{tab1}
\centering 
\begin{tabular}{cc|cccccc}
\toprule
 & &  Mean  & Variance & Kurtosis  & Skewness \\
\midrule
(a) &\textbf{DAX}   &$2.2(2.1) \times 10^{-4}$&$2.2(3) \times 10^{-4}$& 8.3(1.6) & -0.28(14)  \\
&\textbf{Nikkei 225}    &$0.9(1.8)\times 10^{-4}$ &$2.2(3)\times 10^{-4}$&8.6(2.3)& -0.3(1)  \\
&\textbf{SSE}     &$1.3(2.5)\times 10^{-4}$ &$2.2(4)\times 10^{-4}$ &$8.3(9)$ &  $-0.35(21)$ \\
&\textbf{VIX}    &$-0.6(2.7)\times 10^{-4}$ & $4.8(5)\times 10^{-3}$ &9(2) & 0.96(20)    \\
&\textbf{Bitcoin}    & $1.7(1.0) \times 10^{-3}$   &$1.4(2)\times 10^{-3}$& 16(10)  &  -0.80(75) \\
&\textbf{Ethereum}    & $1.6(1.3)\times 10^{-3}$ &$2.6(4)\times 10^{-3}$&$12.5(5.6)$&  $-0.57(63)$   \\

\midrule
(b)&\textbf{DAX}       & $1.03(8)\times 10^{-2}$ & $1.1(2)\times 10^{-4}  $&14(3) & 2.6(3)   \\
&\textbf{Nikkei 225}   & $1.05(7)\times 10^{-2}$ & $1.1(2)\times 10^{-4}  $&17(6) & 2.7(7)  \\
&\textbf{SSE}          & $1.02(9)\times 10^{-2}$ & $1.2(2)\times 10^{-4}  $&13(1) & 2.6(1)  \\
&\textbf{VIX}          & $4.8(2)\times 10^{-2}$  & $2.2(3)\times 10^{-3}  $&20(6) & 2.8(4)   \\
&\textbf{Bitcoin}      & $2.4(3)\times 10^{-2}$  & $7.9(1.9)\times 10^{-4}$&33(24)& 3.4(1.2)   \\
&\textbf{Ethereum}     & $3.4(3)\times 10^{-2}$  & $1.44(23)\times 10^{-4}$&24(14)& $3.1(7)$  \\

\midrule
(c)&\textbf{DAX} &$-0.05(6.82)\times 10^{-4}$&2.6(1)&4.36(18)& $2.7(3.2)\times 10^{-2}$ \\
&\textbf{Nikkei 225}   &$6.4(7.2)\times 10^{-4}$&2.70(7)&4.6(1)& $-3.9(4.5)\times 10^{-2}$  \\
&\textbf{SSE}     & $-4.5(7.3)\times 10^{-4}$ &2.95(11) &4.4(3) & $-0.6(2.6)\times 10^{-2}$ \\
&\textbf{VIX}    &$1.0(2.1)\times 10^{-3}$ &$2.25(7)$ &3.9(1) & $2.6(2.7)\times 10^{-2}$  \\
&\textbf{Bitcoin}    & $0.07(1.86) \times 10^{-3}$& 2.86(7) &4.2(3)&  $3.1(4.0) \times 10^{-2}$  \\
&\textbf{Ethereum}    &$-0.6(1.7)\times 10^{-3}$&2.6(1)&3.58(17)&  $-1.2(5.8)\times 10^{-2}$  \\

\bottomrule
\end{tabular}
\end{table}

\section{Methodology}\label{sec.3}

\subsection{MFDFA}
To examine the multifractal properties of a time series 
we used the MFDFA of \citet{kantelhardt2002multifractal}.
The MFDFA is described using the following steps:

(i) We determine the profile $Y(i)$,
\be
Y(i)=\sum_{j=1}^i (r(j)- \bra r \ket),
\ee
where $\bra r \ket$ denotes the average.

(ii) We divide profile $Y(i)$ into $N_s$ non-overlapping segments of equal length $s$, where $N_s \equiv {int} (N/s)$.
A short period may exist at the end of the profile because the length of the time series is not always a multiple of $s$. 
To utilize this part, the same procedure is repeated, starting from the end of the profile.
Thus, $2N_s$ segments are obtained.

(iii) We calculate the following variance:
\be
F^2(\nu,s)=\frac1s\sum_{i=1}^s (Y[(\nu-1)s+i] -P_\nu (i))^2,
\ee
for each segment, $\nu, \nu=1,\dots,N_s$ and
\be
F^2(\nu,s)=\frac1s\sum_{i=1}^s (Y[N-(\nu-N_s)s+i] -P_\nu (i))^2,
\ee
for each segment, $\nu, \nu=N_s+1,\dots,2N_s$.
where $P_\nu (i)$ is the fitting polynomial used to remove the local trend in segment $\nu$.
\be 
P_\nu (i) =\sum_{k=0}^p a_k i^k.
\ee 
Here, we use a cubic-order $( p=3 )$ polynomial.

(iv) We average over all segments and obtain the $q$th order fluctuation function:
\be
F_q(s)=\left\{\frac1{2N_s} \sum_{\nu=1}^{2N_s} (F^2(\nu,s))^{q/2}\right\}^{1/q}.
\label{eq:FL}
\ee

(v) We determine the GHE $h(q)$ from the scaling exponent of $F_q(s)$. 
If the time series $r(i)$ is long-range power-law correlated,
$F_q(s)$ is expected to have the following functional form for a large $s$:
\be
F_q(s) \sim s^{h(q)}.
\label{eq:asympto}
\ee

The singularity spectrum $f(\alpha)$, which also characterizes the multifractality of the time series,
is defined by $h(q)$ as
\be 
f(\alpha)=q[\alpha-h(q)] +1,
\ee 
where $\alpha$ is the H\"{o}lder exponent or singularity strength, given by
\be 
\alpha= h(q)+qh^\prime(q).
\ee

The Hurst exponent is obtained using $h(2)$.
We restrict the range of $q$ to $q=[-5,5]$ because when $|q|$ is large, the moments in the fluctuation function
diverge, and the calculation of $h(q)$ may be unstable\citep{Jiang-Xie-Zhou-Sornette-2019-RPP}.


We define the strength of multifractality by the width of $h(q)$ as \citep{zunino2008multifractal}
\be 
\Delta h(q)=h(-q)-h(q),
\label{eq1}
\ee
where $q \neq 0$.
Since a Gaussian time series exhibits a monofractal nature with $\Delta h(q) = 0$, a nonzero $\Delta h(q)$ suggests that the strength of multifractality is associated with the degree of market inefficiency.
Similarly, the strength of multifractality can be defined by the width 
of the singularity spectrum $\alpha(q)$ as follows:
\be 
\Delta \alpha(q)=\alpha(-q)-\alpha(q).
\label{eq2}
\ee 
$\Delta \alpha(q)$ is zero for the monofractal time series.

\subsection{Market efficiency measurement by GHE}
Since the Hurst exponent $h(2)$ of a random time series is 0.5, any deviation from this value suggests inefficiencies in the time series.
\citet{wang2009analysis} proposed the market deficiency measure (MDM) defined by 
an average deviation from 0.5 as
\be 
MDM(q)=\frac12 (|h(-q)-0.5|+ |0.5-h(q)|),
\ee 
which is zero in efficient markets.
We observe that $MDM(q)$ reduces $\Delta h(q)$, disregarding a prefactor, for the conditions $h(-q)-0.5>0$ and $0.5-h(q)>0$, which are typically satisfied.
Thus, the multifractal strength $\Delta h(q)$ is also related to inefficiencies quantified by deviations from 0.5.

\section{Empirical results}\label{sec.4}
To obtain the temporal evolution of $h(q)$, we employed the rolling window method and selected an appropriate window size. While smaller window sizes are preferred for better time resolution, statistical fluctuations increase as the window (data) size decreases. Fig. 2 illustrates the time evolution of $h(2)$ for Bitcoin returns, computed using window sizes of 1.5, 2, 3, 4, and 5 years with a rolling window that shifts by one day at a time. We observe that all time evolution patterns exhibit similar trends; however, larger fluctuations are evident with smaller window sizes. Consequently, we selected a 3-year rolling window (1095 days) that demonstrates acceptable fluctuations.
As trading does not occur in the stock market daily,
we selected 750 working days as the approximately 3-year period for the time-series data in the stock market.

\begin{figure}
\centering
\includegraphics[width=11cm]{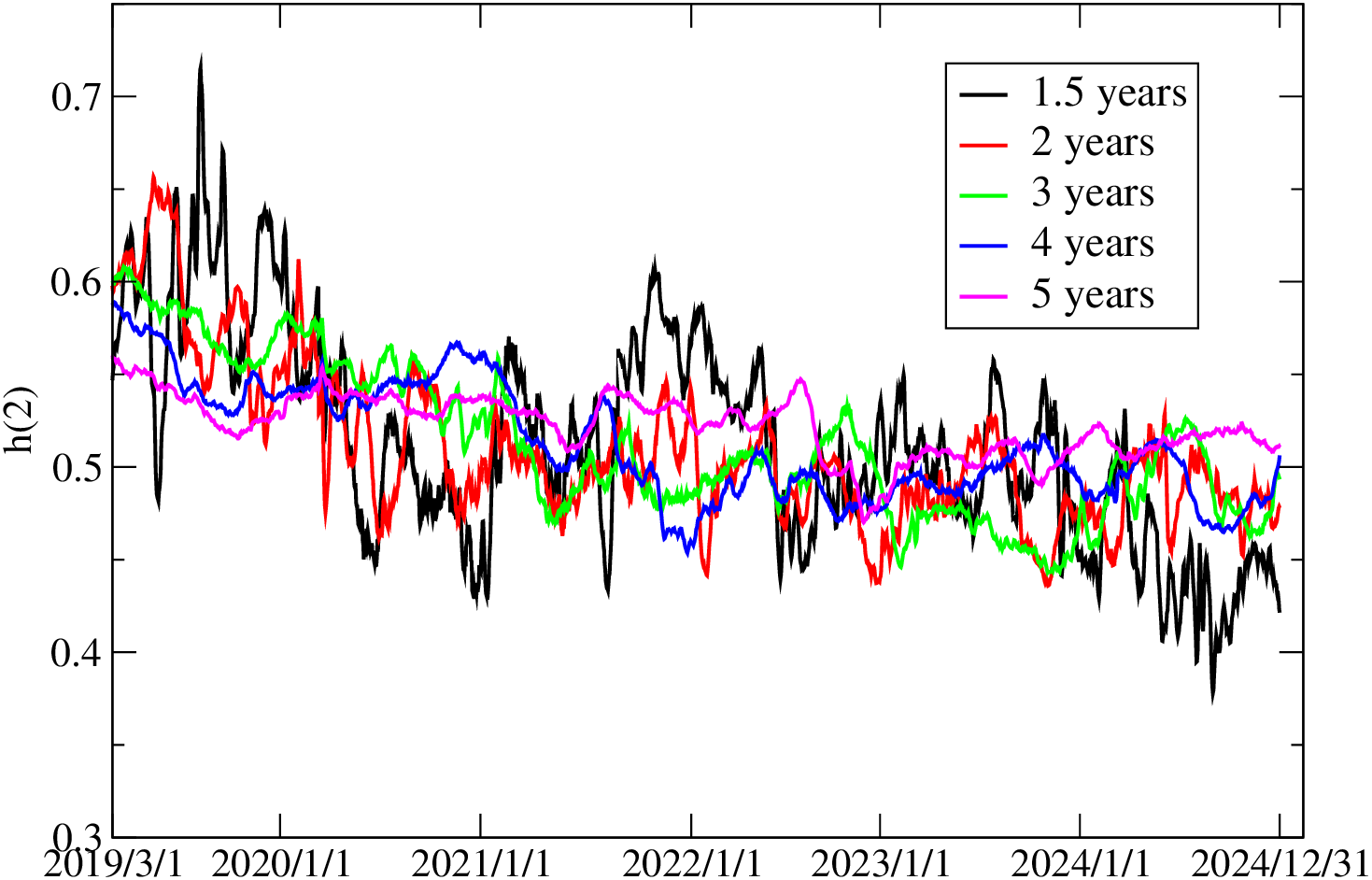}
\caption{Time evolution of $h(2)$ for Bitcoin returns, computed using window sizes of 1.5, 2, 3, 4, and 5 years. }
\end{figure}

First, we show the time evolution of the Hurst exponent $h(2)$ for DAX, Nikkei 225, SSE, and VIX in Fig. 3
and provide an overview of the properties of the four time series.
The Hurst exponents of the returns time series remain approximately 0.5, except for VIX, for which $h(2)$ is less than 0.5, exhibiting anti-persistence.
However, the Hurst exponents of the absolute returns time series 
are greater than 0.5, reflecting the long-memory property of the absolute returns time series \citep{ding1993long}.
The Hurst exponents of the volatility increment time series are small, typically less than 0.1,
which implies that the time series are anti-persistent. 


We then focus on the impact of the Lehman Brothers bankruptcy (the black lines in Fig. 3).
We find no clear response to the impact of the Lehman Brothers bankruptcy in the
time evolution of $h(2)$ except in the case of DAX and Nikkei 225, where it increased promptly. 
No clear impact of the Lehman Brothers’ bankruptcy on other stocks and the results of returns and volatility increments was observed.

\begin{figure}[]
\centering
\subfloat[\centering]{\includegraphics[width=5.8cm]{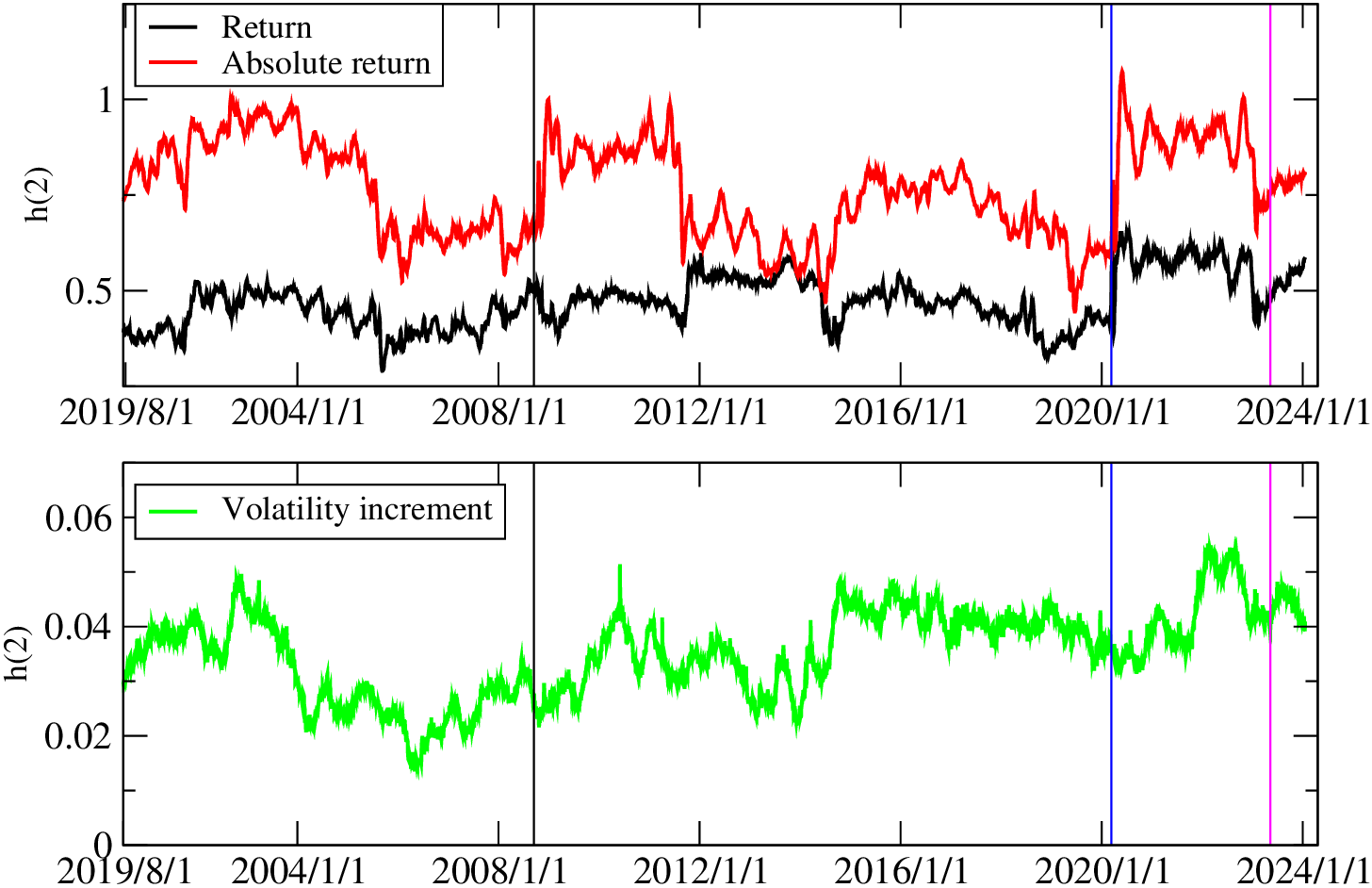}}
\hfill
\subfloat[\centering]{\includegraphics[width=5.8cm]{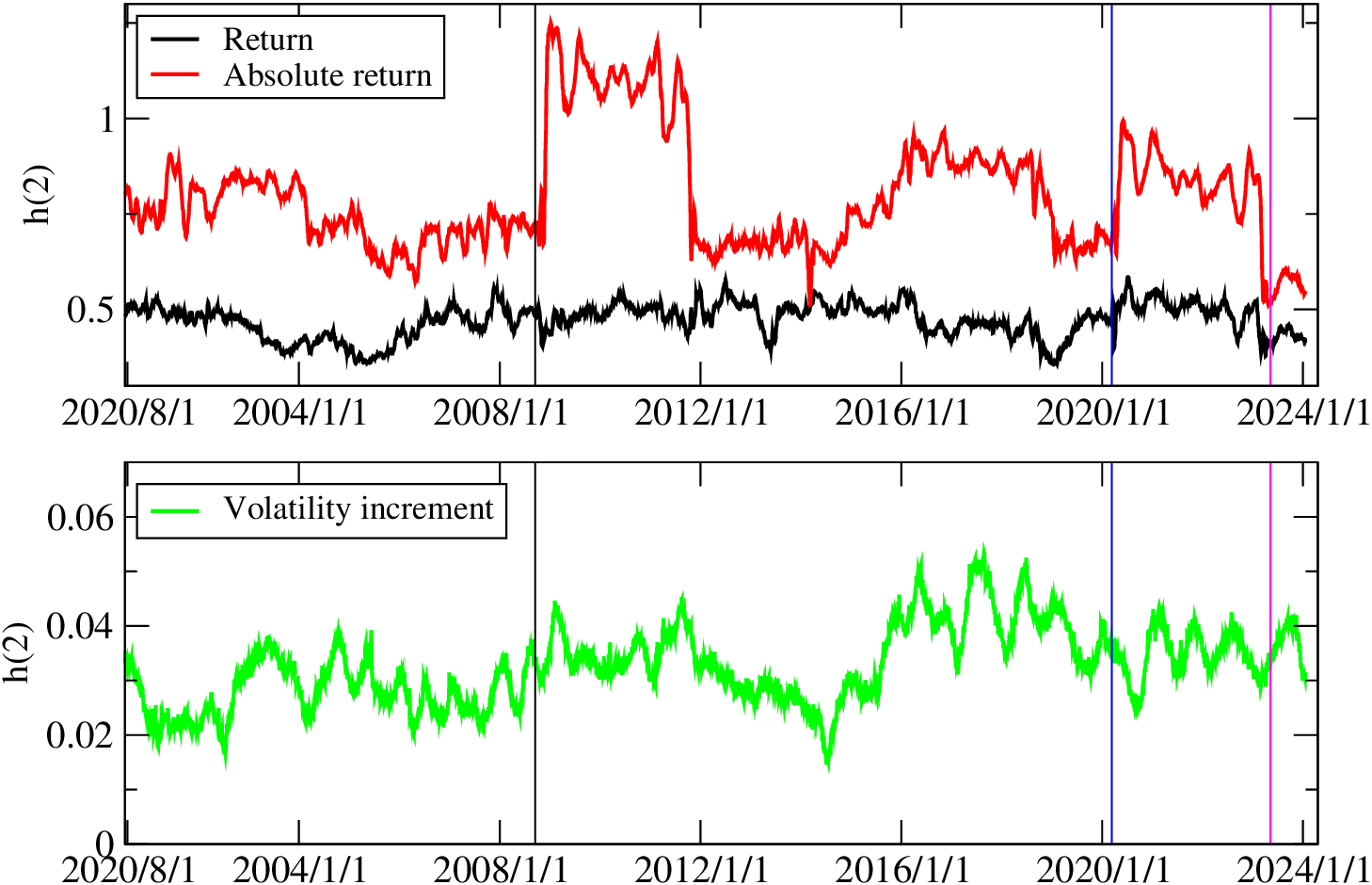}}\\
\subfloat[\centering]{\includegraphics[width=5.8cm]{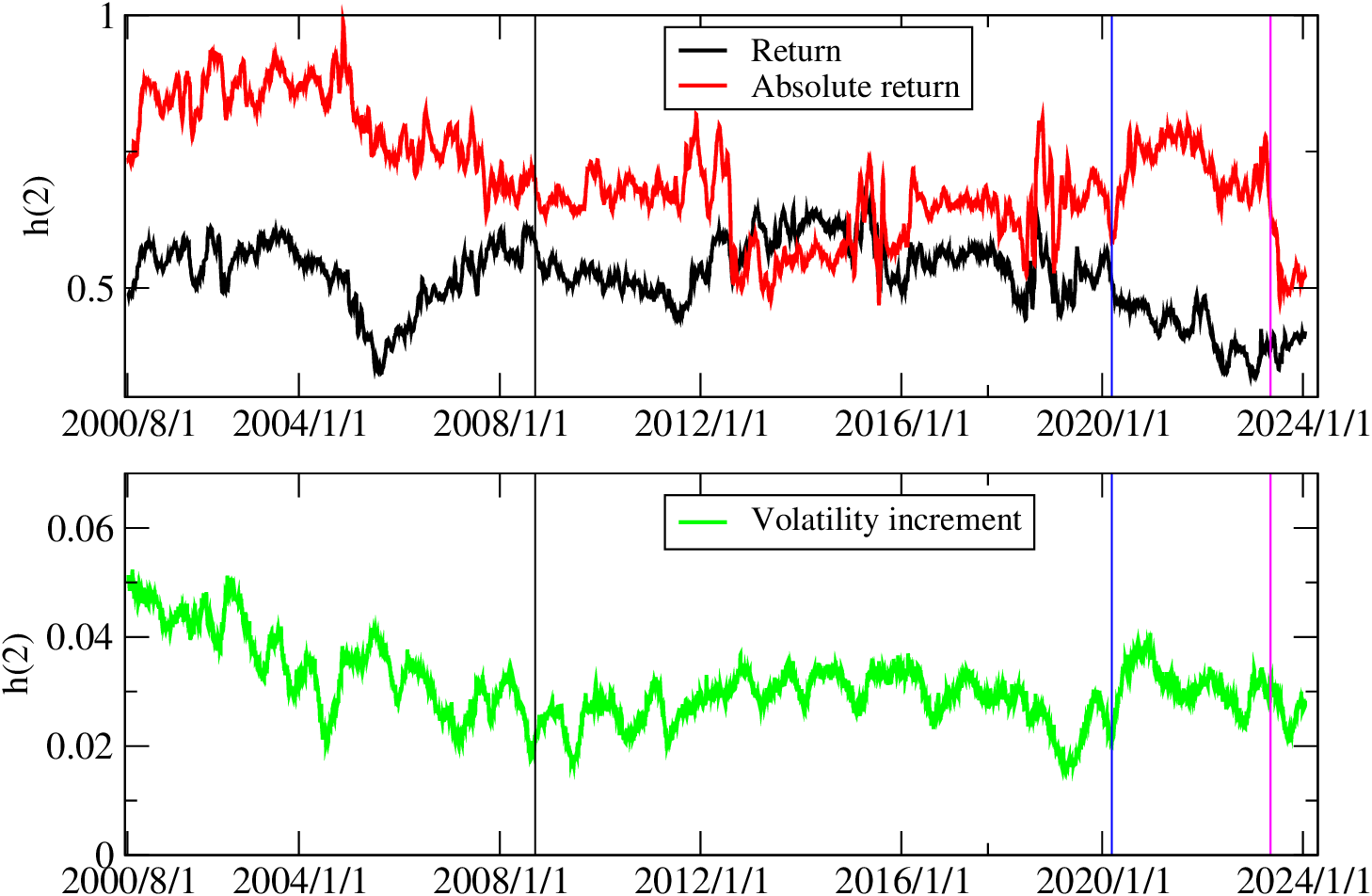}}
\hfill
\subfloat[\centering]{\includegraphics[width=5.8cm]{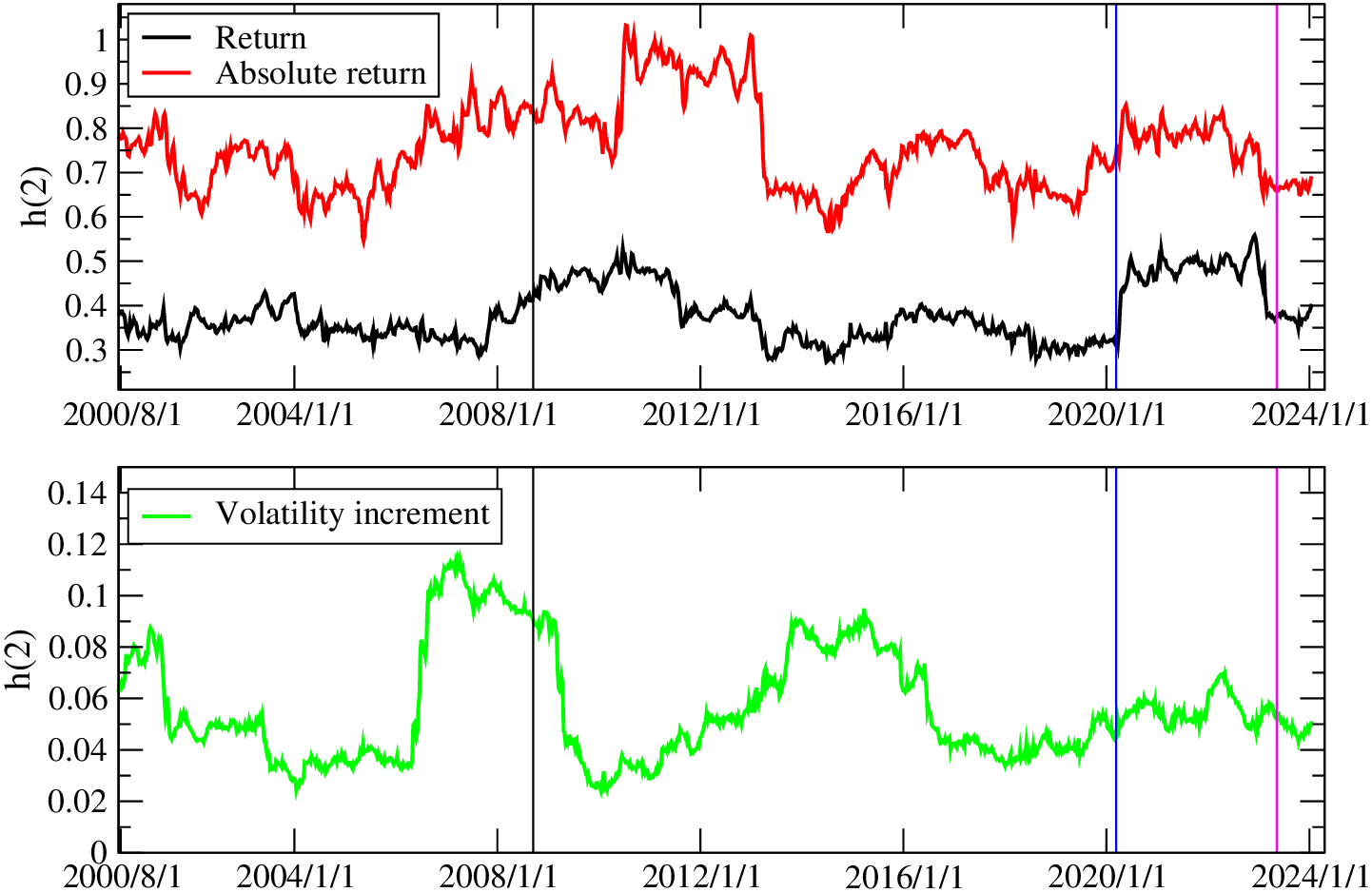}}
\caption{Time evolution of $h(2)$ for returns, absolute returns, and volatility increment time series in the stock market. (\textbf{a}) DAX, (\textbf{b}) Nikkei 225, (\textbf{c}) SSE, (\textbf{d}) VIX.
The three lines in the plots indicate the financial and pandemic-related date 
as follows.
 (\textbf{black}) The bankruptcy of Lehman Brothers (2008/9/15). (\textbf{blue}) The WHO declared COVID-19 a pandemic (2020/3/11). (\textbf{magenta}) The WHO declared the end of the COVID-19 pandemic (2023/5/5).
\label{fig2}}
\end{figure} 

To examine the COVID-19 period in more detail, 
we plotted the results within the period from July 1, 2019, to January 19, 2024, in Fig. 4.
We find that the pandemic had a definite impact on both returns and absolute returns.
However, no impact of the pandemic was observed for the $h(2)$ of the volatility increment.
In the case of DAX and Nikkei 225, the $h(2)$ of returns and absolute returns reacted sensitively to 
the WHO declaration of a pandemic on March 11, 2020. Around the day of the WHO declaration,
we observe that $h(2)$ increased abruptly, then suddenly decreased, and then increased again.
During the pandemic, $h(2)$ remained higher than before the pandemic.
After the WHO's declaration at the end of the pandemic, $h(2)$ returned to its previous levels. 

Regarding the SSE and VIX, we observed a reaction similar to that of DAX and Nikkei 225,
that is, the $h(2)$ of the returns on VIX and the $h(2)$ of the absolute returns on the SSE and VIX 
increased around the time of the pandemic declaration and returned to previous levels after the declaration at the end of the pandemic.
An exceptional case is $h(2)$ for the returns of the SSE that did not react to the pandemic.

\begin{figure}[]
\centering
\subfloat[\centering]{\includegraphics[width=5.7cm]{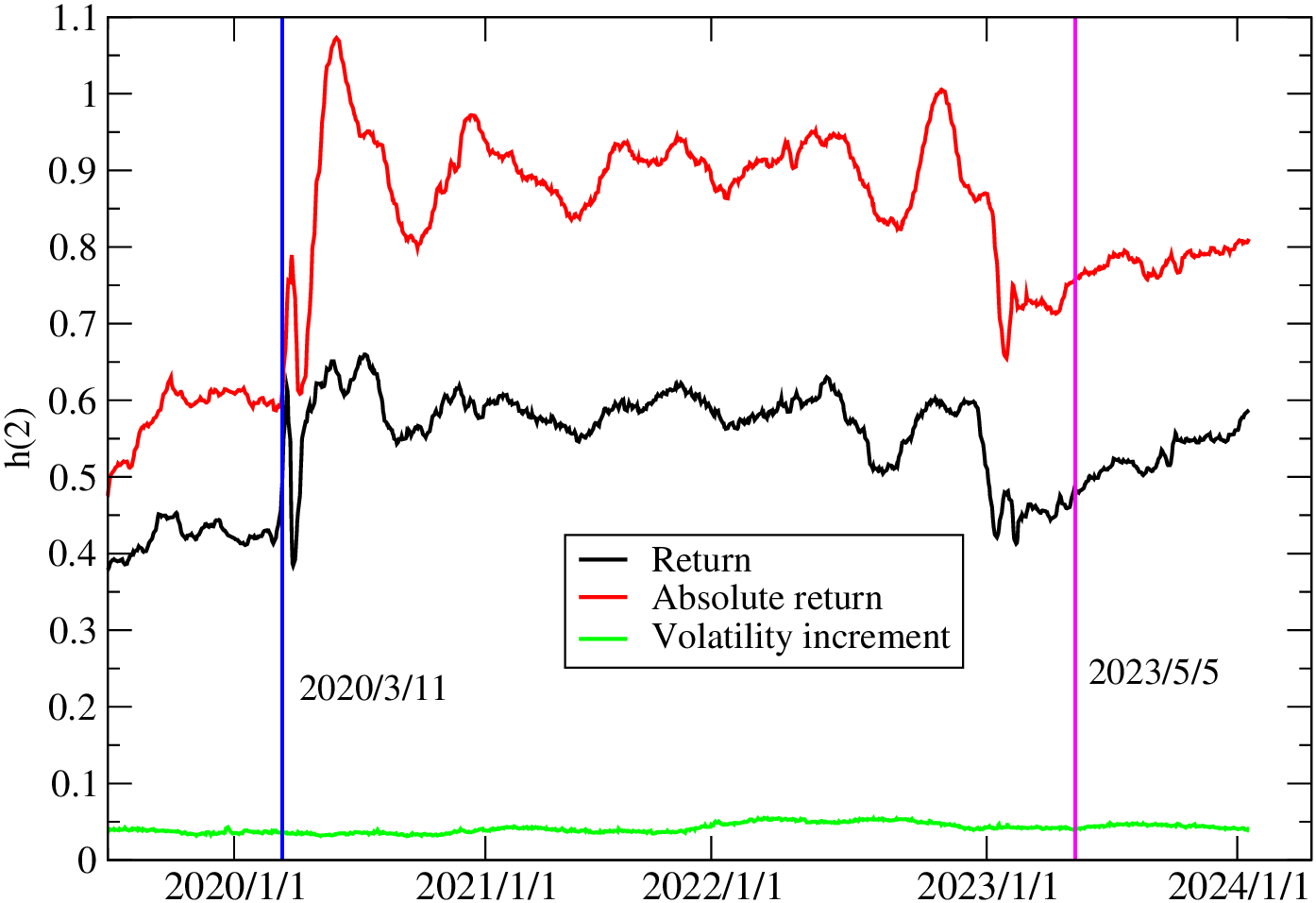}}
\hfill
\subfloat[\centering]{\includegraphics[width=5.7cm]{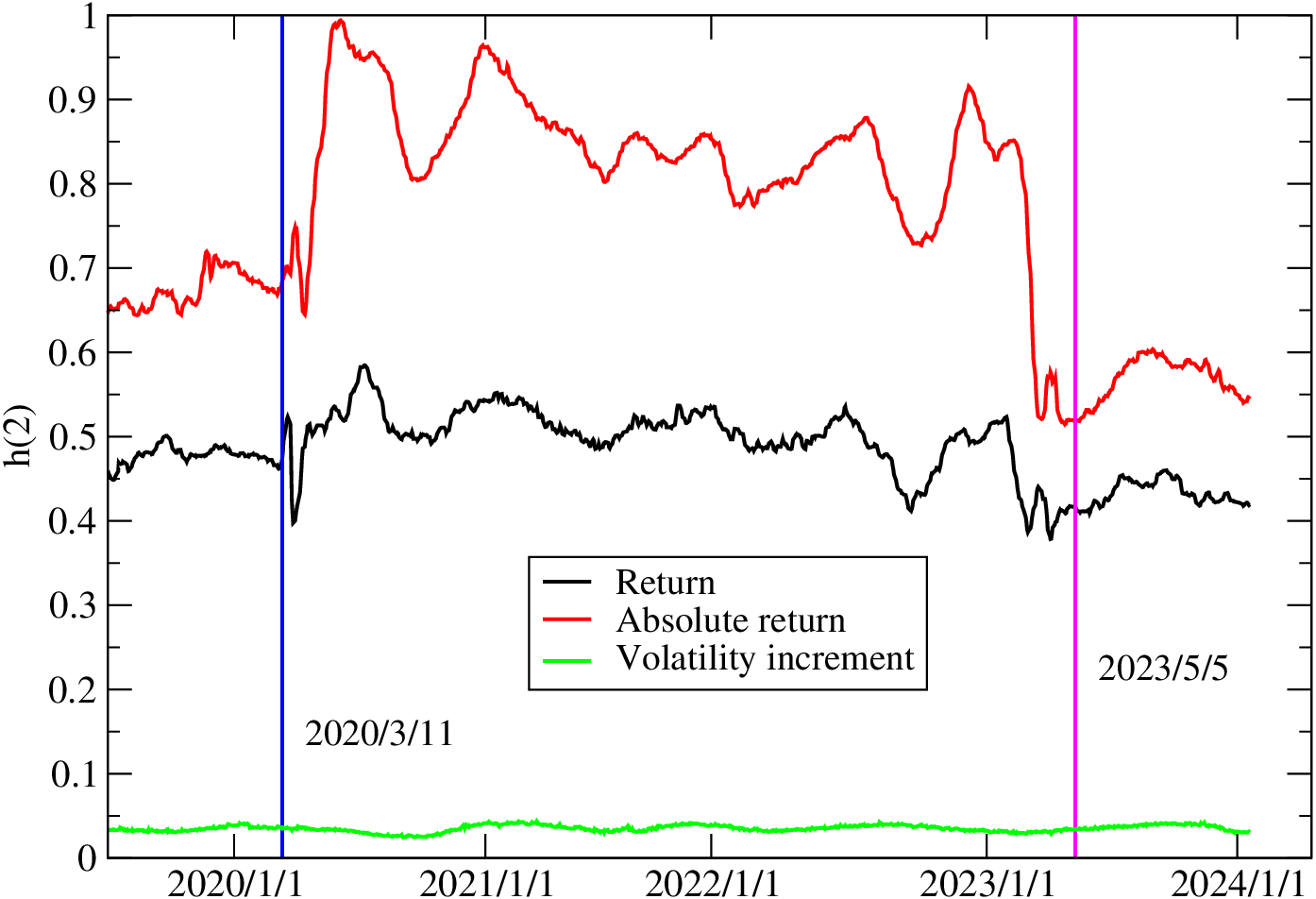}}\\
\subfloat[\centering]{\includegraphics[width=5.7cm]{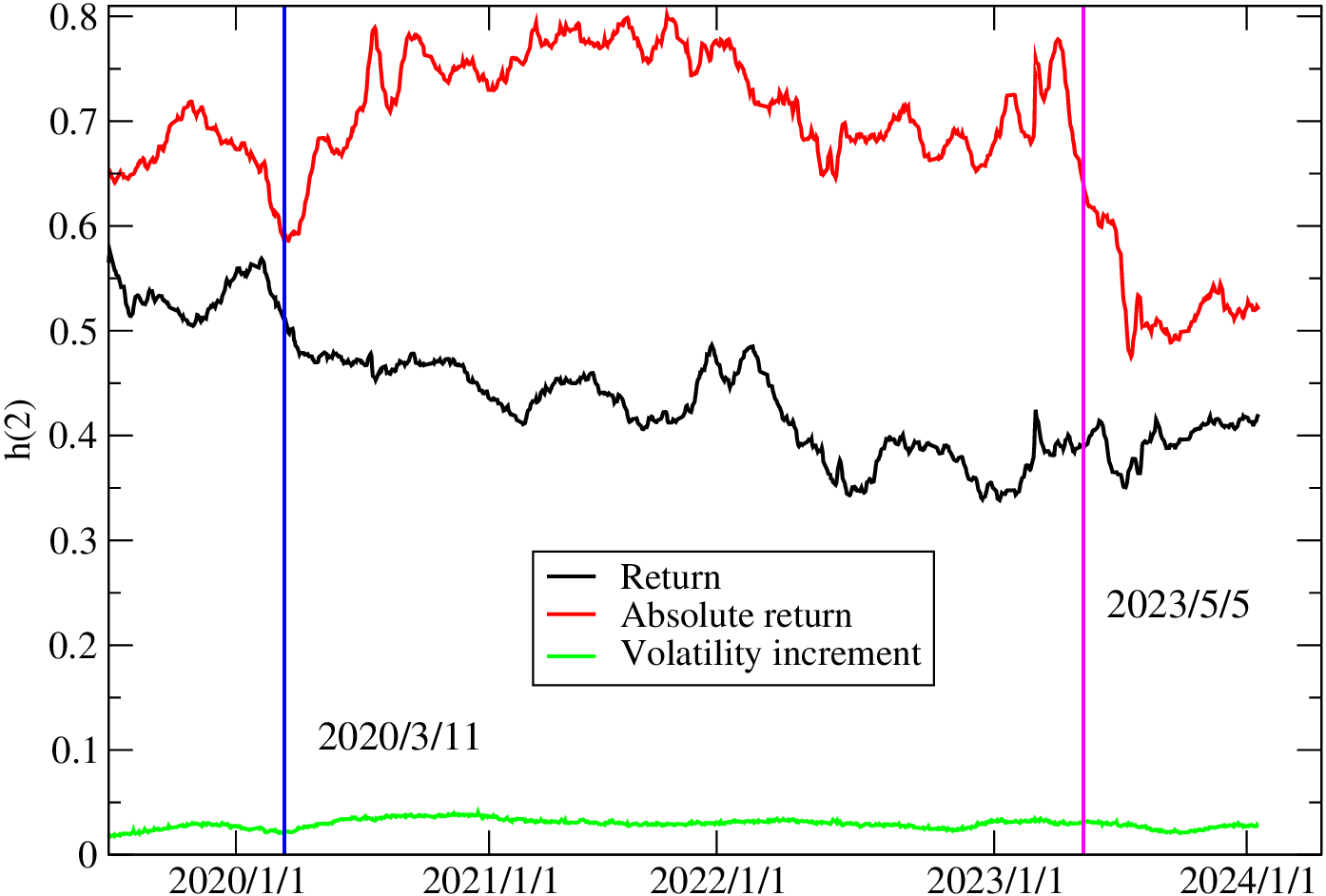}}
\hfill
\subfloat[\centering]{\includegraphics[width=5.7cm]{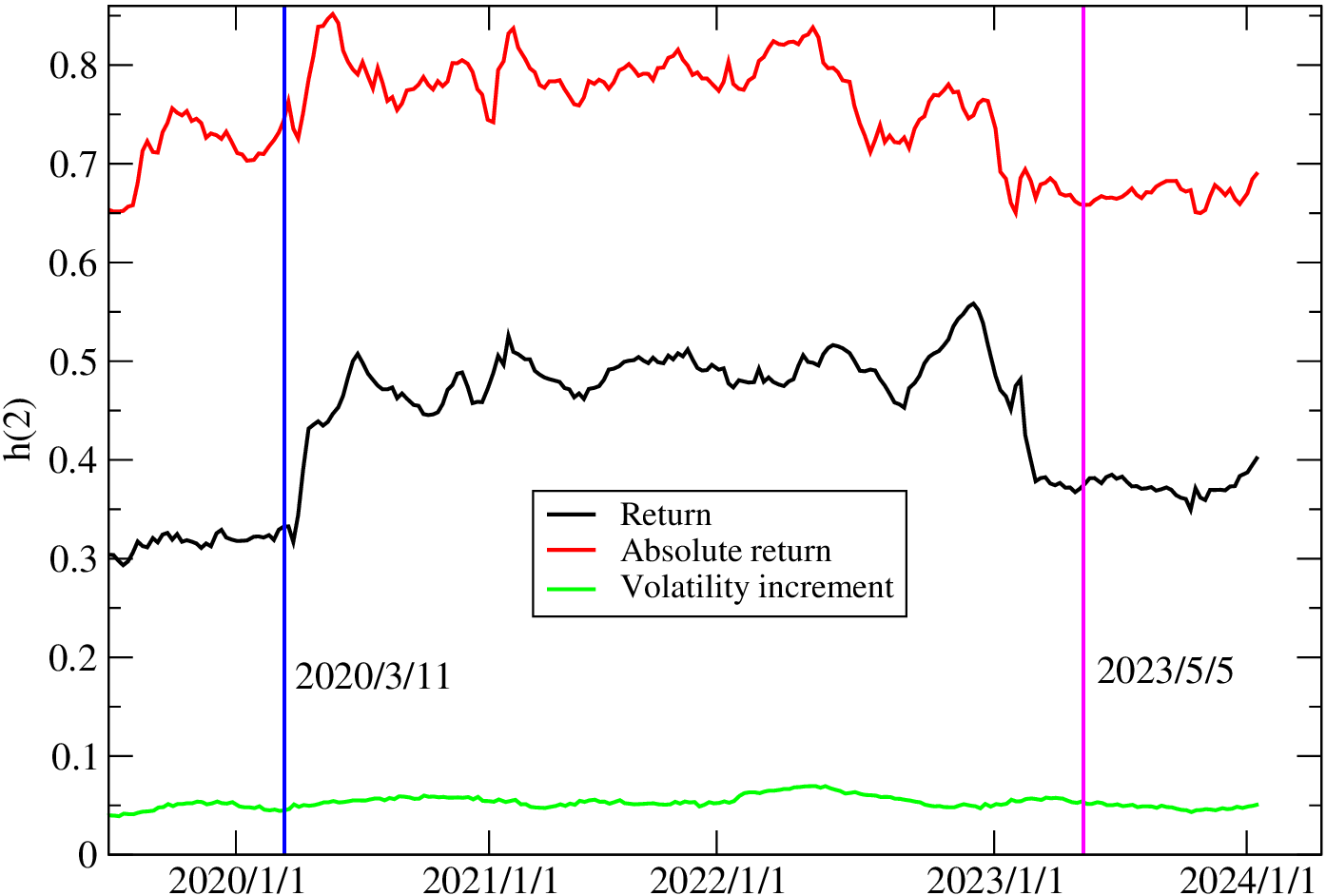}}
\caption{
Time evolution of $h(2)$ for returns, absolute returns, and volatility increment time series in the stock market, from July 1, 2019 to January 19, 2024.
(\textbf{a}) DAX, (\textbf{b}) Nikkei 225, (\textbf{c}) SSE, (\textbf{d}) VIX.
The two dotted lines indicate the pandemic-related date as follows.
(\textbf{blue}) The WHO declared COVID-19 a pandemic (2020/3/11). (\textbf{magenta}) The WHO declared the end of the COVID-19 pandemic (2023/5/5).
}
\end{figure}

\begin{figure}[]
\centering
\subfloat[\centering]{\includegraphics[width=9cm]{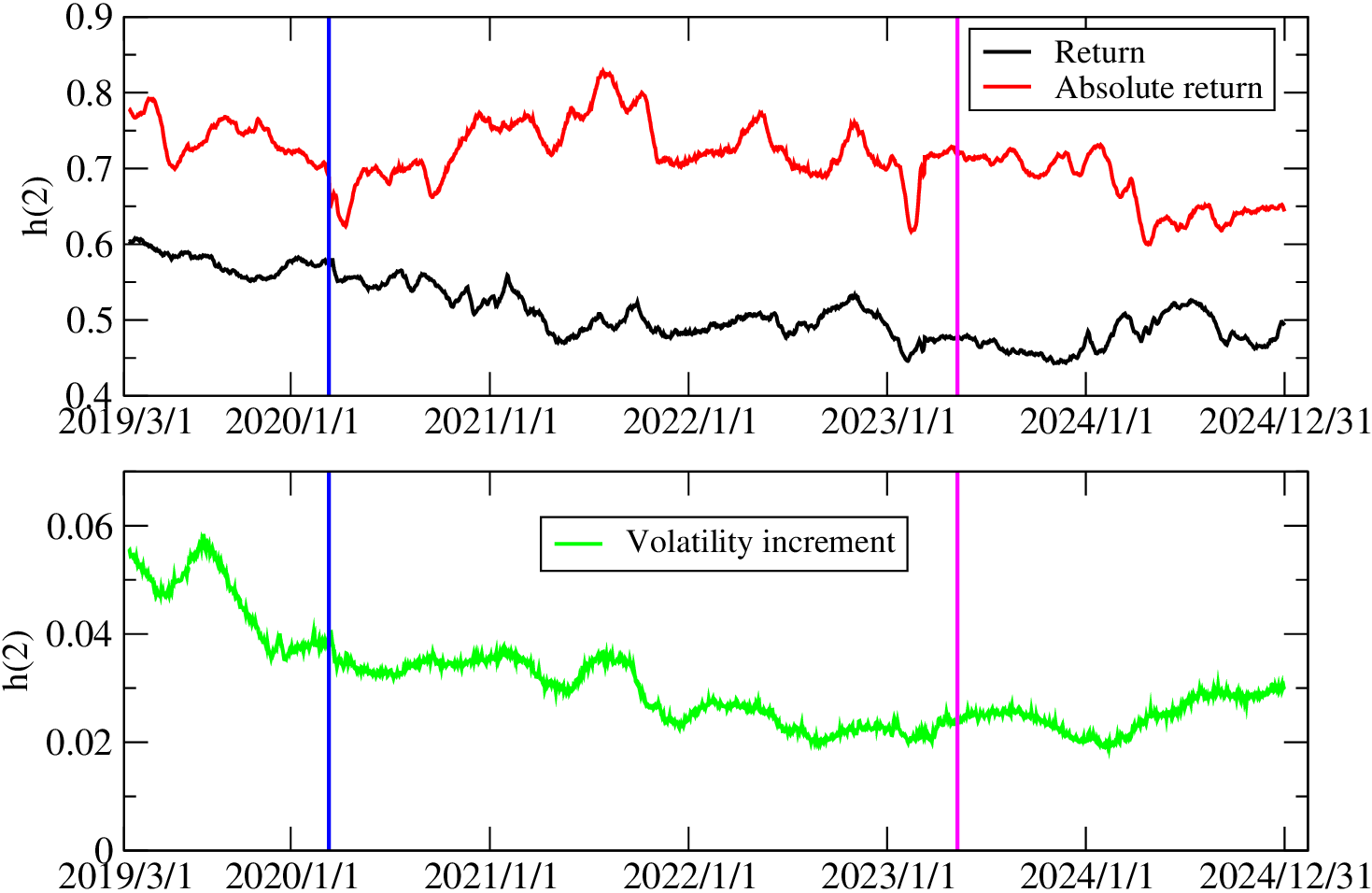}}
\hfill
\subfloat[\centering]{\includegraphics[width=9cm]{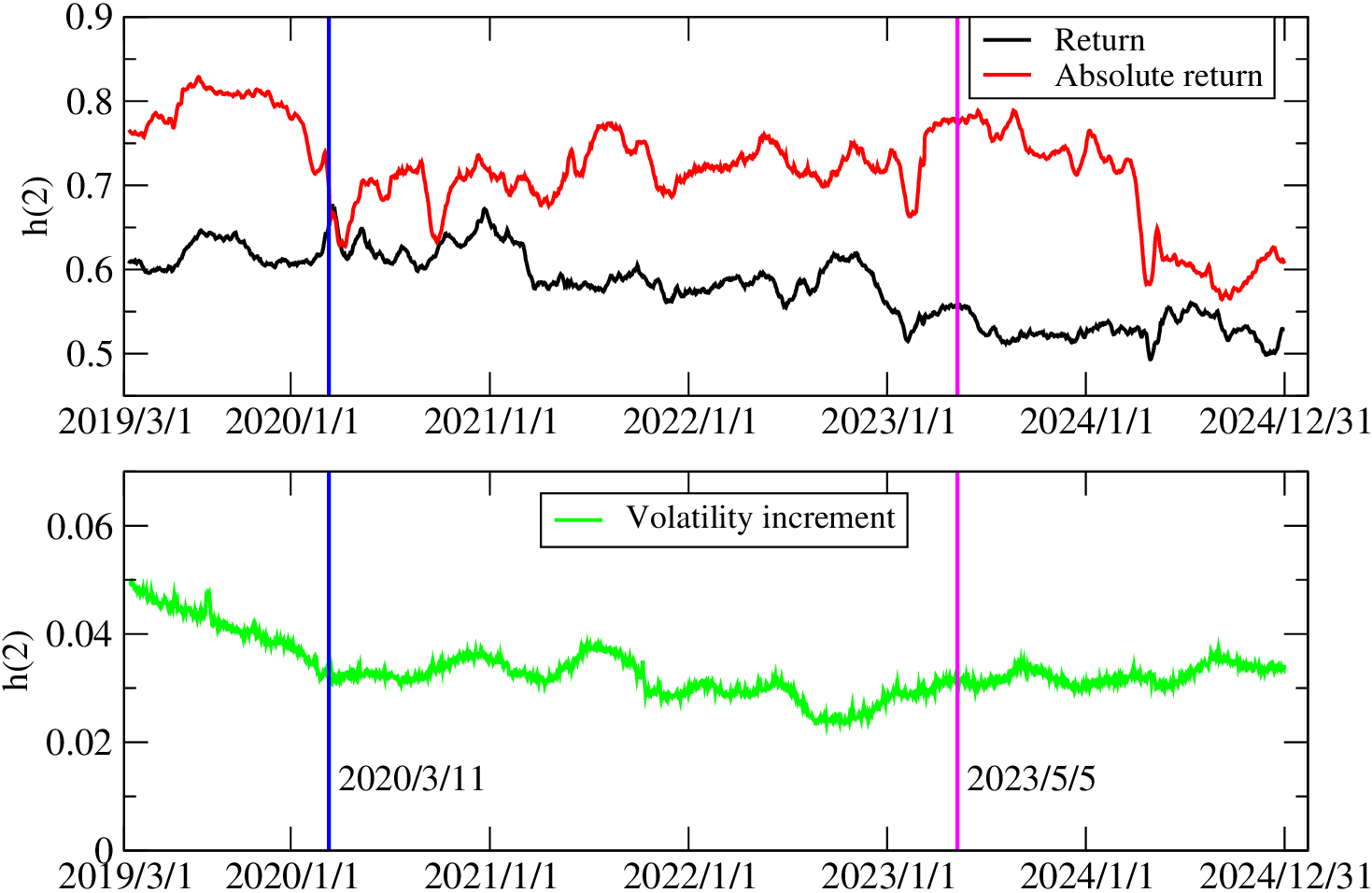}}\\
\caption{
Time evolution of $h(2)$ for returns, absolute returns, and volatility increment time series in the cryptocurrency market. (\textbf{a}) Bitcoin, (\textbf{b}) Ethereum.
The two lines indicate the pandemic-related date as follows.
(\textbf{blue}) The WHO declared COVID-19 a pandemic (2020/3/11). (\textbf{magenta}) The WHO declared the end of the COVID-19 pandemic (2023/5/5).}
\end{figure}

Fig. 5 shows the results of $h(2)$ for the cryptocurrencies Bitcoin and Ethereum.
These results suggest that there was no discernible response to the pandemic. 
In contrast to the stock market, the $h(2)$ of cryptocurrencies exhibits no reaction to the pandemic.
Although the $h(2)$ of returns for Bitcoin and Ethereum appeared to decrease gradually, we cannot conclude that this gradual decline is attributable to the pandemic.
If we conduct a period-based analysis in this case, we might observe a pseudo-pandemic effect.
For example, Table 2 shows the $h(2)$ of returns for Bitcoin and Ethereum before and during the COVID-19 periods, 
indicating that as the Hurst exponent gradually decreased, market efficiency appeared to improve after the COVID-19 pandemic.

\begin{table}
\centering
\caption{Hurst exponents of returns for Bitcoin and Ethereum before and during the COVID-19 periods.}
\label{tab1}
\centering 
\setlength{\tabcolsep}{3.8mm}\begin{tabular}{c|cc}
\toprule
 period &  Bitcoin  & Ethereum   \\
\midrule
January 1, 2017--December 31, 2019 (before COVID-19) &  0.578 & 0.605 \\
January 1, 2020--December 30, 2020 (during COVID-19) &   0.493 &  0.561\\
\bottomrule
\end{tabular}
\end{table}

Similar to the stock market, the $h(2)$ of the volatility increment is low and exhibits anti-persistence.
This anti-persistence is consistent with that observed for the realized and implied volatility time series, although the Hurst exponents of the volatility increment defined by 
absolute returns lead to a smaller Hurst exponents\citep{gatheral2018volatility}. 
For example, the Hurst exponent of Bitcoin realized volatility is calculated as approximately 0.12-14\citep{takaishi2025multifractality}.
However, the Bitcoin volatility increment from absolute returns volatility 
was approximately 0.02-0.04 in recent years, as shown in Fig. 5, which is smaller than that of Bitcoin realized volatility.
This reduction can be explained by the finite sample effect on the realized volatility.
Realized volatility is constructed using a finite sample of squared intraday returns \citep{andersen1998answering,mcaleer2008realized}, and
its accuracy depends on sample size \citep{peters2006testing,andersen2007no,fleming2003economic}. \citet{takaishi2025multifractality} found that the Hurst exponent decreases 
as the sample size decreases and provides a phenomenological formula that describes
the Hurst exponent as a function of sample size.
\be
H_2(n)=H_2\frac{n}{n+a_1},
\label{eq3}
\ee 
where $n$ is the sample size and $a_1$ is a fitting parameter.
$H_2$ denotes the Hurst exponent at $n \rightarrow \infty$.
The absolute returns (or their square) corresponds to the lowest-order realized volatility with $n=1$.
Using $a_1\sim 3$ obtained in \citet{takaishi2025multifractality} and $n=1$, 
we obtain
\be 
H_2 =H_2(1)\times 4.
\ee 
Fig. 6 shows the Hurst exponent of Bitcoin volatility increment 
and its values multiplied by four.  
The results multiplied by four are close to the Hurst exponent obtained for Bitcoin realized volatility ($0.12\sim0.14$), 
except for the results ($\sim 0.2$) around 2019.

\begin{figure}
\centering
\includegraphics[width=11cm]{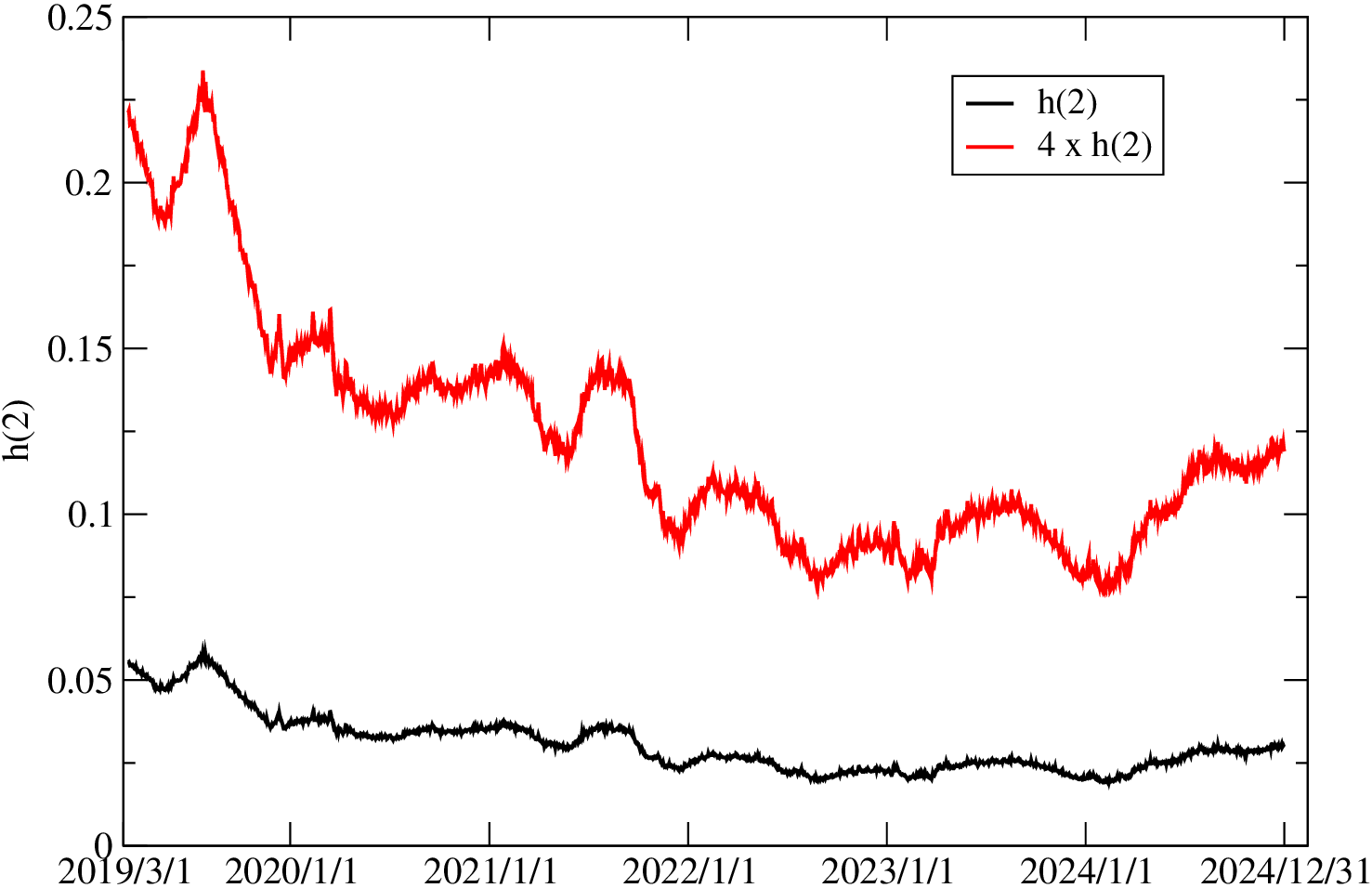}
\caption{The Hurst exponent $h(2)$ and its values multiplied by four. \label{fig2}}
\end{figure}

Finally, we analyzed the effect of the pandemic on multifractal strength $\Delta h(n)$.
Although $\Delta h(n)$ depends on $n$, we observe that the temporal variation pattern exhibits similar fluctuations regardless of $n$, as illustrated in Fig. 7. Consequently, by selecting $n=5$, we focus on $\Delta h(5)$ in the subsequent analysis.
\begin{figure}
\centering
\includegraphics[width=11cm]{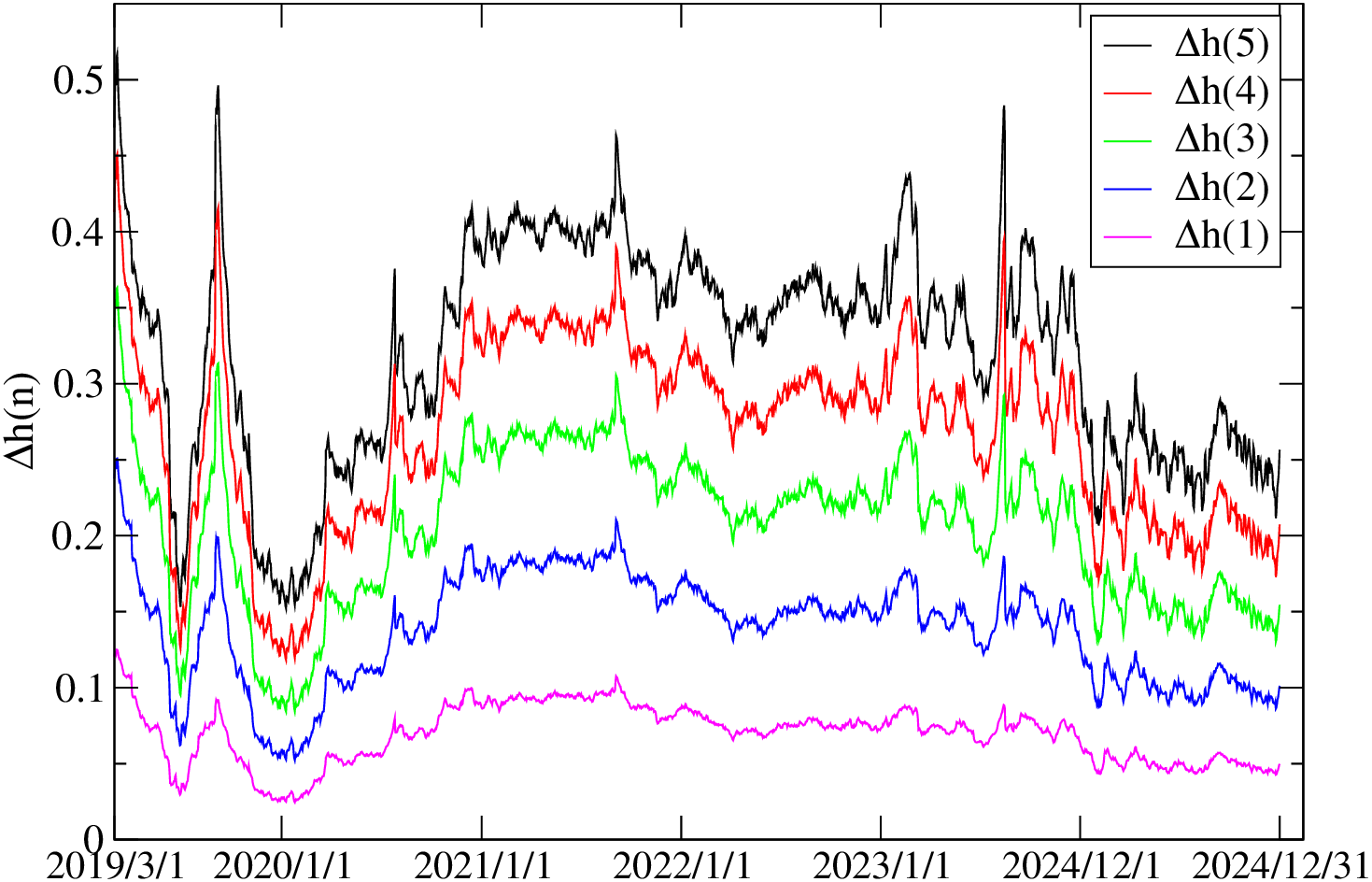}
\caption{Time evolution of $\Delta h(n)$ for Bitcoin returns, computed for $n=1,2,3,4$ and $5$.}
\end{figure}  

The stock market results of $\Delta h(5)$ are shown in Fig. 8.
We observe that the $\Delta h(5)$ of the volatility increment remained stable with positive values and
did not respond to the pandemic.
However, the $\Delta h(5)$ of the returns and absolute returns fluctuated dramatically, and
negative values can be assumed. 
This observation implies that the functional form of $h(q)$ changed significantly over time. 
Although the $\Delta h(5)$ of the returns and absolute returns for DAX and Nikkei 225 
appeared to have responded to the pandemic, it is difficult to identify a pattern describing the impact of the pandemic.

\begin{figure}[]
\centering
\subfloat[\centering]{\includegraphics[width=5.7cm]{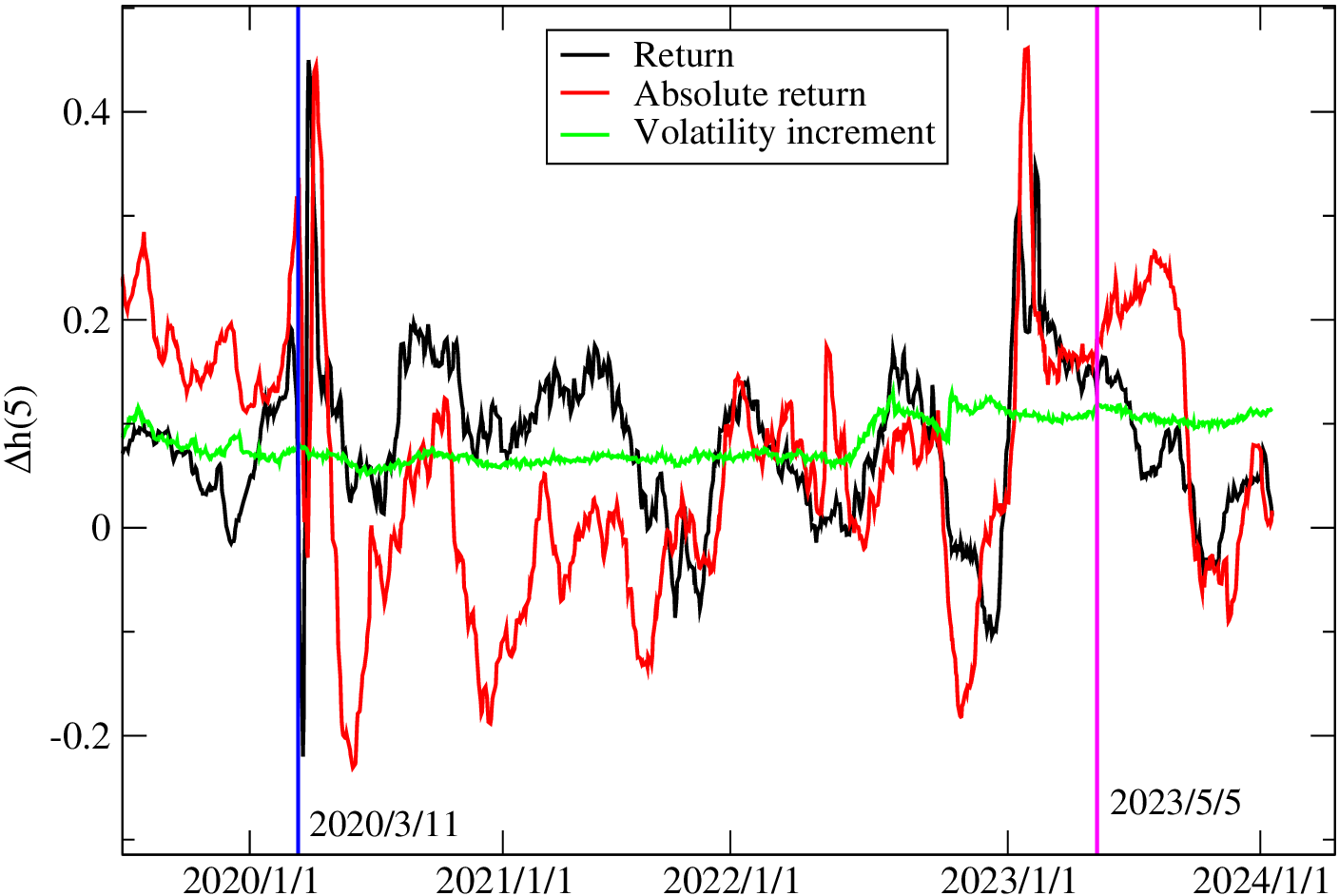}}
\hfill
\subfloat[\centering]{\includegraphics[width=5.7cm]{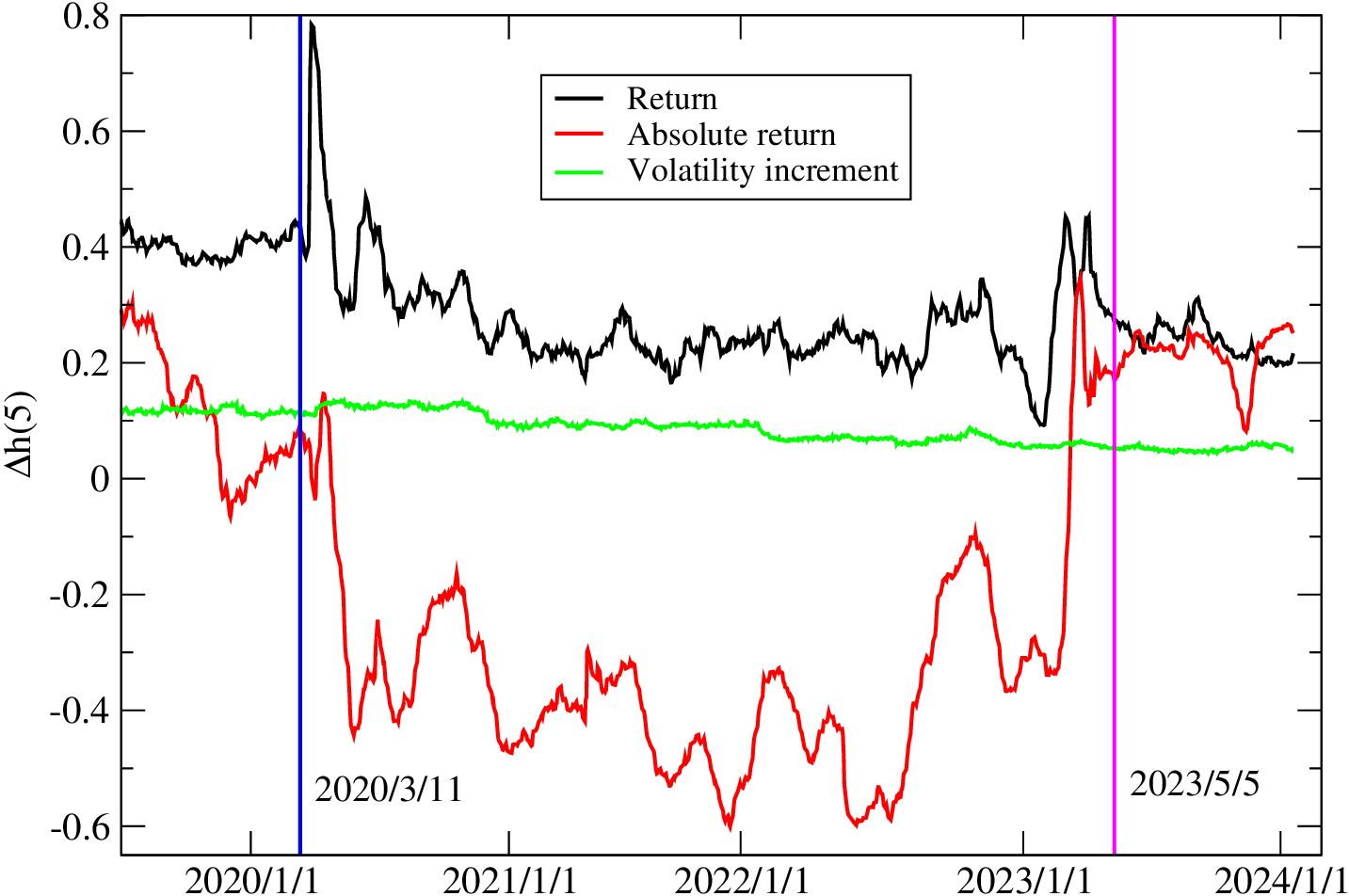}}\\
\subfloat[\centering]{\includegraphics[width=5.7cm]{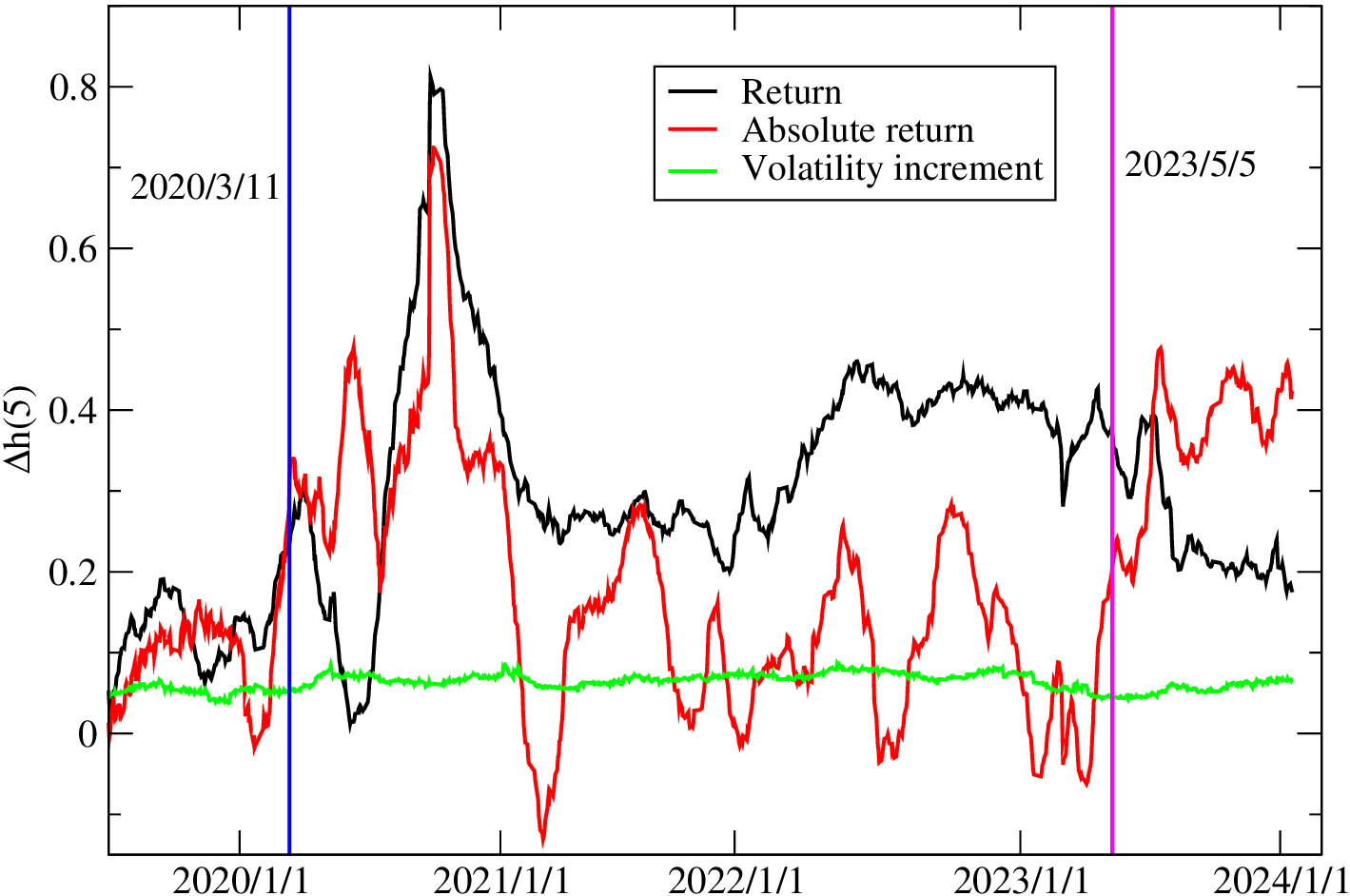}}
\hfill
\subfloat[\centering]{\includegraphics[width=5.7cm]{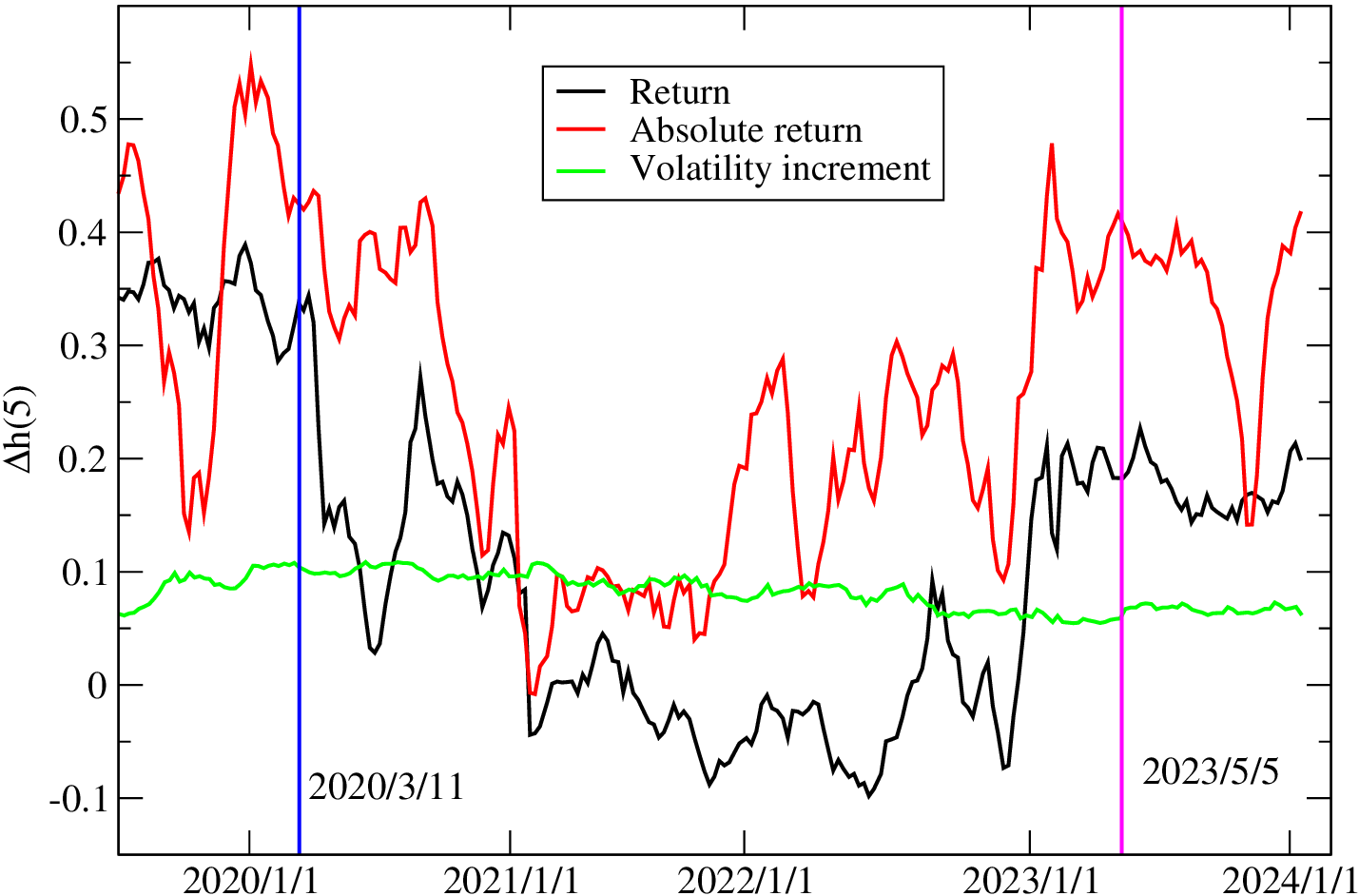}}
\caption{
Same as Fig.4 but for time evolution of $\Delta h(5)$.
\label{fig6}}
\end{figure}

Fig. 9 shows the time evolution of $\Delta h(5)$ for cryptocurrencies.
Similar to the stock market, the $\Delta h(5)$ of the volatility increment for cryptocurrencies remained stable with positive values.
Although $\Delta h(5)$ values of the returns and absolute returns fluctuate considerably,
we identify a pattern in which $\Delta h(5)$ increased around the time of the pandemic declaration and
obtained a smaller $\Delta h(5)$ value after the end of the COVID-19 pandemic.

\begin{figure}[]
\centering
\subfloat[\centering]{\includegraphics[width=6cm]{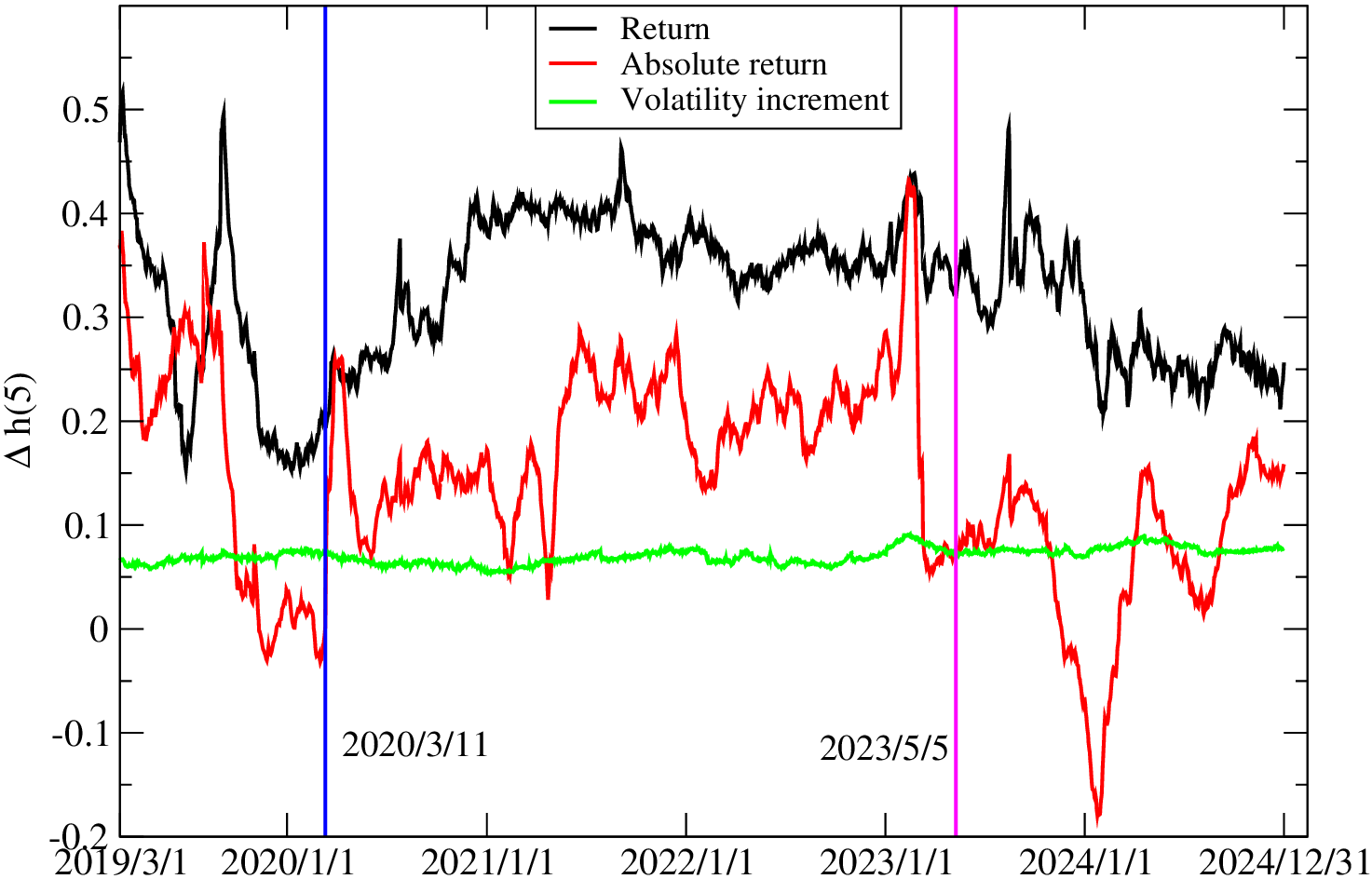}}
\hfill
\subfloat[\centering]{\includegraphics[width=6cm]{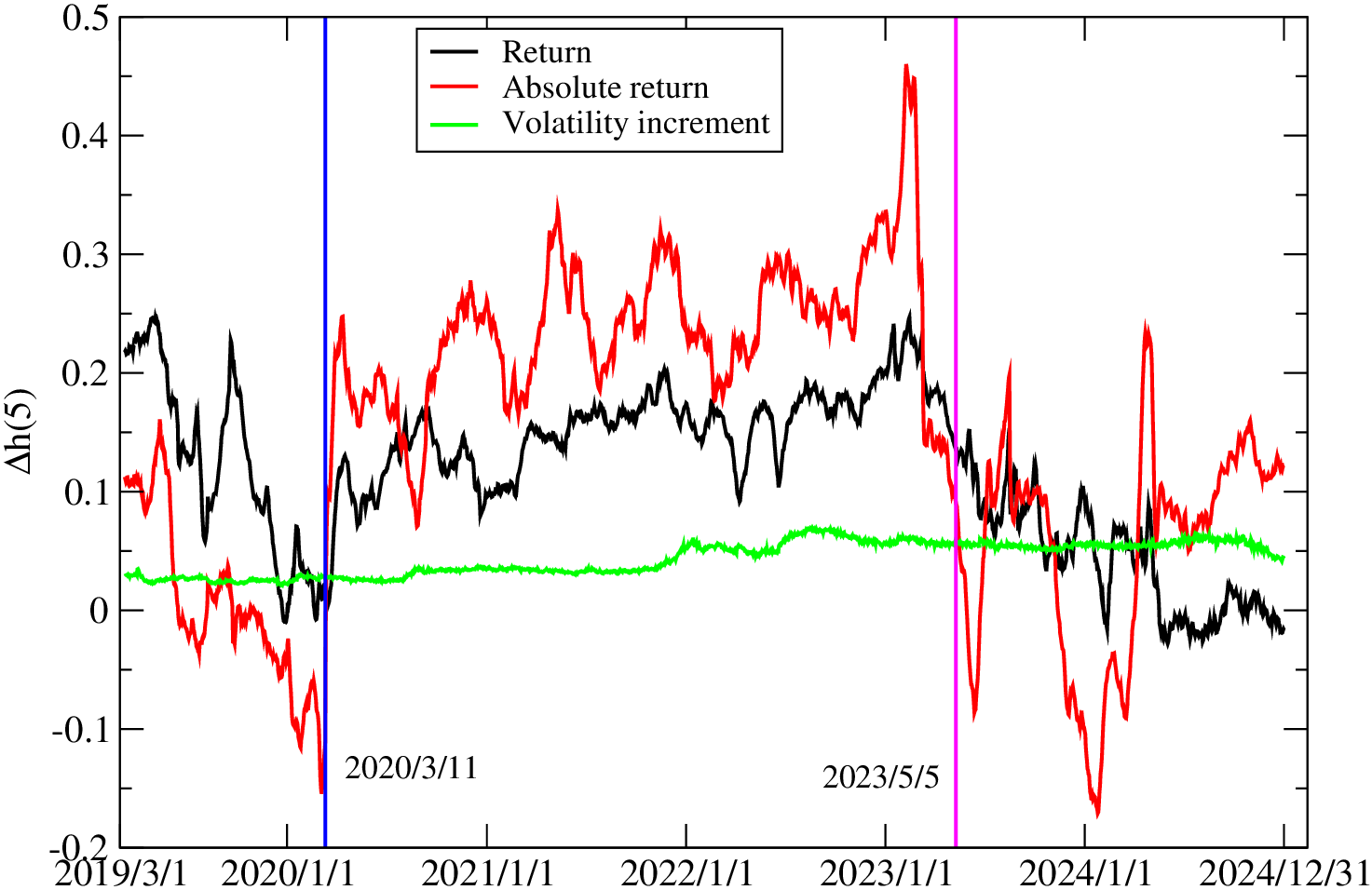}}\\
\caption{
Same as Fig.5 but for time evolution of $\Delta h(5)$.
}
\end{figure}

The multifractal strength is also defined by the singularity spectrum $\alpha(q)$ as in Eq.(\ref{eq2}).
We compared $\Delta \alpha(5)$ with $\Delta h(5)$
and discovered that both exhibit similar time-variation patterns.
For example, Fig. 10 displays the time evolution of $\Delta \alpha(5)$ and $\Delta h(5)$ for Bitcoin, demonstrating similar temporal patterns. 
Therefore, in such cases, 
the same conclusion can be drawn regarding the temporal variations.
A comparison of $\Delta \alpha(q)$ and $\Delta h(q)$ for Bitcoin returns is also provided in \citep{takaishi2021time}.

\begin{figure}[]
\centering
\subfloat[\centering]{\includegraphics[width=5.7cm]{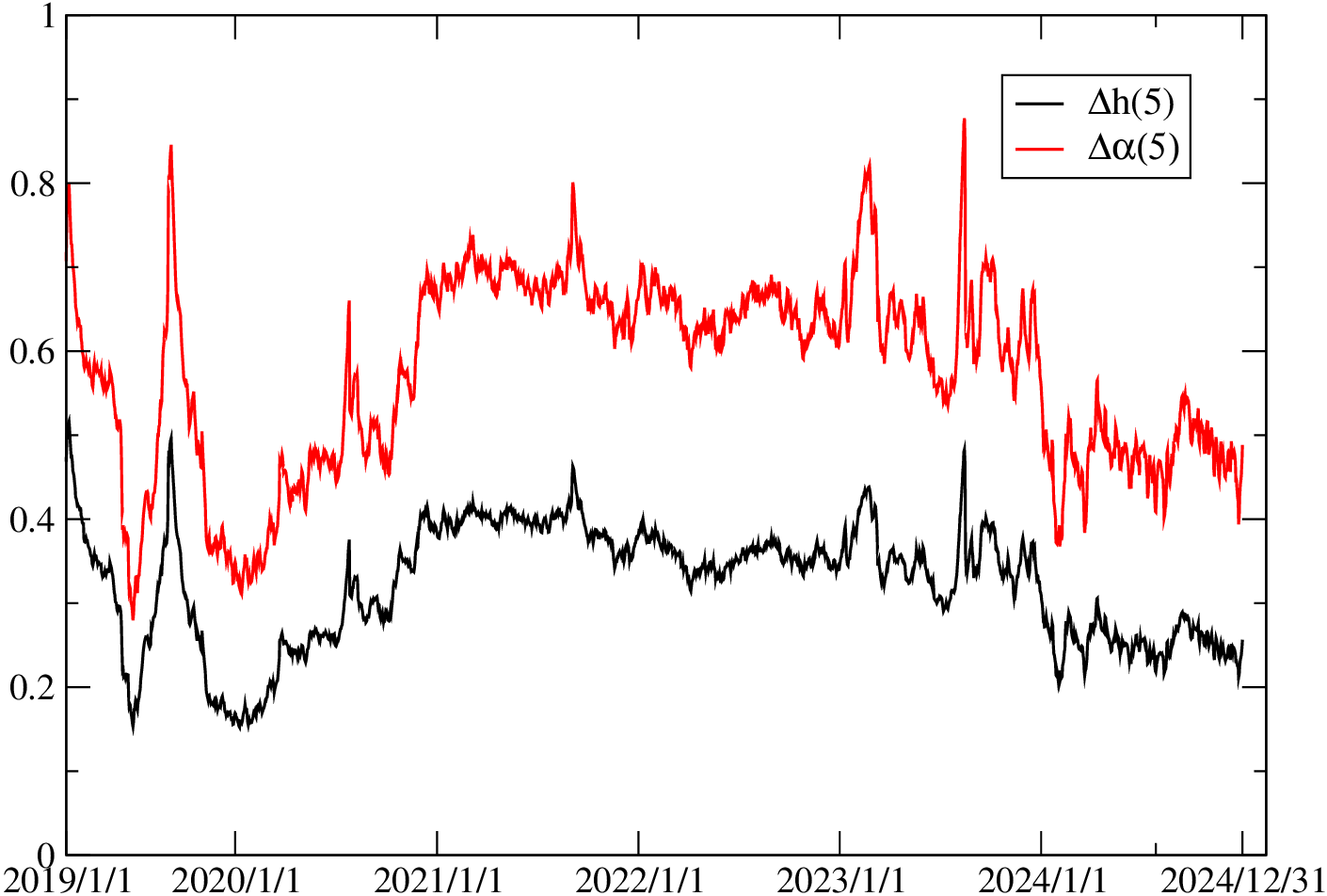}}
\hfill
\subfloat[\centering]{\includegraphics[width=5.7cm]{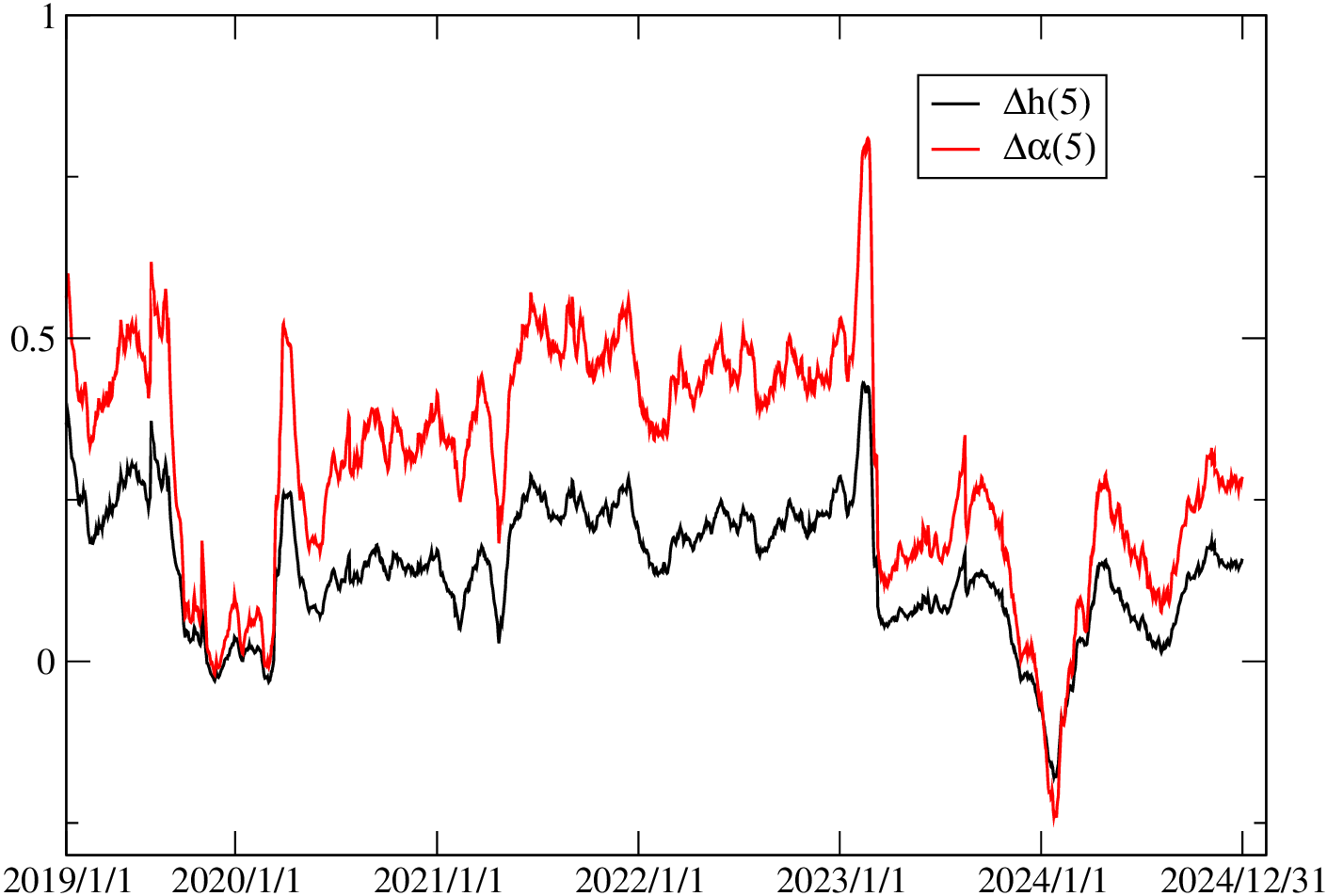}}\\
\subfloat[\centering]{\includegraphics[width=5.7cm]{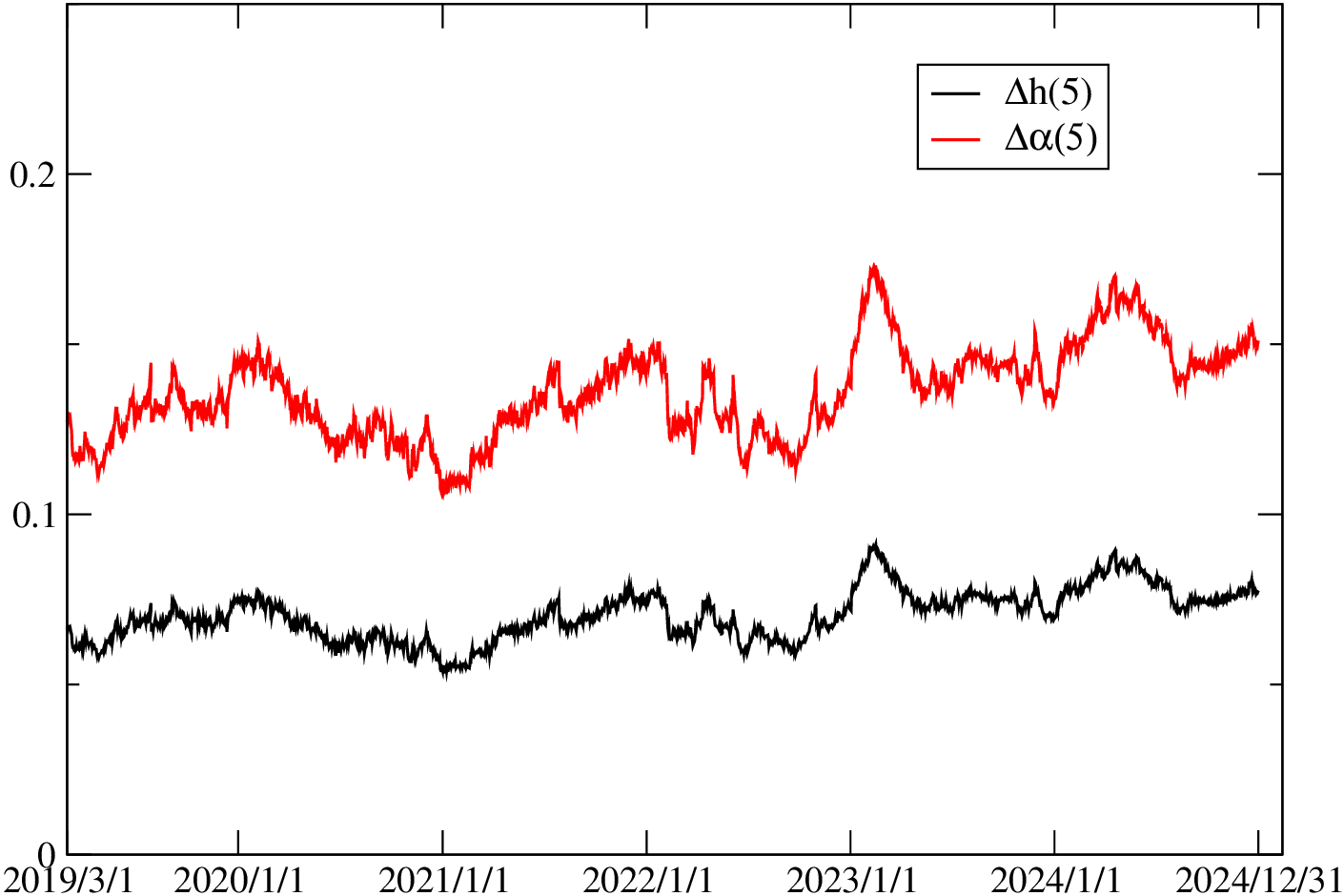}}
\hfill
\caption{Comparison of $\Delta h(5)$ and $\Delta \alpha(5)$ for Bitcoin. 
(\textbf{a}) Returns, (\textbf{b}) absolute returns, (\textbf{c}) volatility increment.
}
\end{figure} 

Table 3 summarizes whether the pandemic affected the time series.
When the pandemic has an impact, its effect on efficiency is classified as ``positive'' if it leads to improvement and ``negative'' if it has the opposite outcome.
When an impact has been observed but improvements in efficiency are unclear, it is represented by the symbol $\bigcirc$.
The table indicates that the pandemic exerted varying impacts depending on the differences in markets and types of time series.
Broadly speaking, the COVID-19 pandemic has affected all markets. However, a notable feature of the stock market is that the Hurst exponent of returns for SSE shows no apparent impact, which may indicate that the Chinese market possesses distinct characteristics compared to other markets.
Similarly, in the cryptocurrency market, the Hurst exponent $h(2)$ of returns for Bitcoin and Ethereum did not exhibit significant changes, further suggesting unique characteristics relative to other markets. In contrast, the Hurst exponent $h(2)$ of returns for VIX has been affected in a way that enhances market efficiency, reinforcing the notion that it has distinct characteristics compared to other markets.

\begin{table}
\centering
\caption{The COVID-19 pandemic's impact on returns, absolute returns (AR), and volatility increment (VI) time series. The meanings of the symbols are as follows. $\bigcirc$: impact observed,  $\times$: no impact observed,  $\triangle$: impact unclear.  }
\label{tab1}
\centering 
\setlength{\tabcolsep}{3.8mm}\begin{tabular}{c|cccccc}
\toprule
  &  \textbf{DAX}  & \textbf{Nikkei 225} & \textbf{SSE} & \textbf{VIX} & \textbf{Bitcoin} & \textbf{Ethereum} \\
\midrule
Returns $h(2)$         &negative   &$\bigcirc$ &$\times$     & positive  & $\times$    & $\times$  \\
Returns $\Delta h(5)$  &$\bigcirc$ &$\bigcirc$ & negative    & positive  &  negative   &  negative \\
AR $h(2)$             &negative   &negative   & negative    & negative  & $\times$    & $\times$  \\
AR  $\Delta h(5)$     &$\bigcirc$ &negative   & $\triangle$ & positive  & $\triangle$  & negative \\
VI $h(2)$             &$\times$   &$\times$   &$\times$     & $\times$  & $\times$  & $\times$\\
VI $\Delta h(5)$      &$\times$   &$\times$   &$\times$     & $\times$  & $\times$  & $\times$ \\

\bottomrule
\end{tabular}
\end{table}

\section{Discussion and Conclusions}\label{sec.5}
We investigated the impact of the COVID-19 pandemic on the time evolution of market efficiency for returns, absolute returns, and volatility increments in the stock and cryptocurrency markets.
In the case of stock markets, we found that the $h(2)$ of returns and absolute returns responded to the COVID-19 pandemic 
except in the case of SSE (representative of the Chinese market). We observed no response to the pandemic for the $h(2)$ of returns for the SSE,
which is consistent with previous observations that the Chinese market reacted to the pandemic differently \citep{ameer2023impact,saadaoui2023skewed}.
However, an increased response to the COVID-19 pandemic
was observed for the absolute returns on the SSE.  
Thus, market efficiency depends on the time series used, and 
market efficiency must be viewed from multiple perspectives.

We found that the Hurst exponent $h(2)$ of VIX returns was predominantly below 0.5, indicating the anti-persistence of the time series. This observation aligns with the claim of \citet{bariviera2023disentangling} that anti-persistence arises due to the mean-reverting nature of VIX.

Although we observed no response to the COVID-19 pandemic in the $h(2)$ of the absolute returns for Bitcoin and Ethereum, \citet{arouxet2022covid} reported a temporary reaction in the $h(2)$ of absolute returns. \citet{arouxet2022covid} utilized high-frequency data, such as 5-minute returns, and daily returns may lack sufficient resolution to capture such transient effects.
This argument also extends to the volatility increment time series. While the volatility increment time series appears unaffected by the pandemic, high-frequency data may reveal short-term variations.

The multifractal strength $\Delta h(5)$ of the returns and absolute returns for the stock indices fluctuates,
and it is difficult to identify the effect of the COVID-19 pandemic using $\Delta h(5)$.
In the case of Bitcoin and Ethereum, the $\Delta h(5)$ of the returns and absolute returns increased 
after the declaration of the COVID-19 pandemic. Although an increase in multifractal strength is often used as a sign of a reduction in efficiency, caution is required in interpreting efficiency using multifractal strength, as multifractality is induced by broad distributions \citep{kantelhardt2002multifractal}. Thus, careful analysis---such as that of \citet{li2023dynamic,gao2024impact,choi2021analysis}---may be required to distinguish between the intrinsic and distributional effects 
for the multifractal strength.
\citet{drozdz2020complexity} observed a heavier-tailed returns distribution, which may have induced stronger multifractality during the pandemic.

Several previous studies have examined the effects before and after the COVID-19 pandemic.
However, such analyses do not always reveal the impacts of a pandemic.
For example, we found that the $h(2)$ of returns for SSE, Bitcoin, and Ethereum gradually decreases over time. 
Therefore, when the calculations were divided into periods, the latter may yield smaller values, which do not necessarily clarify whether the pandemic had an impact.

An increase in market inefficiency suggests the presence of arbitrage opportunities. Therefore, understanding the impact of the pandemic on market efficiency provides valuable insights into trading strategies for traders and investors. 
Furthermore, as the pandemic may have had adverse effects on markets, accurately quantifying its impact is crucial for policymakers to stabilize markets.
Our findings indicate that the impact varies depending on the type of market and time series, highlighting the necessity of assessing data across diverse markets and time series categories. For instance, the Hurst exponent in the cryptocurrency market showed no influence from the pandemic, unlike the stock market, suggesting fundamental differences in market characteristics. 
Moreover, our results indicated that the period-based method may not always accurately capture the impact of the pandemic. This underscores the importance of continuously monitoring market efficiency.

The pandemic also impacted risk management methods. As mentioned above, the pandemic may have led to a heavy-tailed distribution of returns which, in turn, could have affected risk measures such as Value at Risk (VaR) and Conditional VaR (CVaR). Therefore, investigating the relationship between multifractality, which is associated with distributional shape, and risk measures like VaR and CVaR would be of significant interest.

Numerous studies on market efficiency have utilized the MFDFA methodology, which has the distinct advantage of being applicable to non-stationary time series. Nonetheless, to ensure the robustness of the findings, it may be prudent to investigate alternative methods, such as entropy-based, wavelet-based, and structure function techniques, which have already been applied in GHE analyses \citep{gatheral2018volatility,lahmiri2023effect,arouxet2022covid}.



\section*{Acknowledgments}
The numerical calculations for this study were performed using
Yukawa Institute Computer Facility and the Institute of Statistical Mathematics.
This research was funded by a Grant-in-Aid from the Zengin Foundation for Studies on Economics and Finance and, in part, by JSPS KAKENHI Grant Number [JP21K01435]




\bibliographystyle{elsarticle-harv} 


\end{document}